\documentclass[journal]{IEEEtran}
\IEEEoverridecommandlockouts
\usepackage[utf8]{inputenc}
\usepackage[T1]{fontenc}
\usepackage{graphicx}
\usepackage{epsfig} 
\usepackage[bottom]{footmisc}
\usepackage{cite}
\usepackage{amsmath, amssymb, amsfonts}
\usepackage{fancyhdr}
\usepackage{float}
\usepackage{algorithm}
\usepackage{algpseudocode} 
\usepackage{algorithmicx}
\usepackage{textcomp}
\usepackage{multirow}
\usepackage[xcdraw, table]{xcolor}
\usepackage{booktabs} 
\usepackage{caption}
\usepackage{subcaption}
\usepackage{rotating}
\usepackage{tablefootnote}
\usepackage{tikz}
\usepackage{ctable} 
\def\BibTeX{{\rm B\kern-.05em{\sc i\kern-.025em b}\kern-.08em
    T\kern-.1667em\lower.7ex\hbox{E}\kern-.125emX}}
\begin{document}

\title{Model-Agnostic Uncertainty Quantification for Fast NFC Tag Identification using RF Fingerprinting
}

\author{Dickson Akuoko Sarpong,~Adam Kamrath,~Rohit Bhusal,~and~Hongzhi~Guo,~\IEEEmembership{Senior Member,~IEEE}  
\thanks{The authors are with the University of Nebraska-Lincoln, Lincoln, NE, 68588, USA. E-mails: dakuokosarpong2@huskers.unl.edu, akamrath2@huskers.unl.edu, rbhusal2@huskers.unl.edu and hguo10@unl.edu. }
\thanks{This work was supported in part by National Science Foundation under grant no. CNS-2341846 and CNS-2310856.}
}

\maketitle

\begin{abstract}
Near Field Communication (NFC) is widely used in security applications such as door access systems and ID cards. However, clone attacks can replicate digital information, enabling unauthorized access. RF fingerprinting offers a robust defense by extracting unique physical-layer features from NFC cards that cannot be cloned. While RF fingerprinting has been extensively applied to Internet of Things (IoT) device authentication, NFC tags present distinct characteristics that require specialized approaches.
This paper focuses on RF fingerprinting for the ISO15693 NFC tag, which is a widely used international standard, by leveraging multi-channel, multi-rate data sampling to enhance accuracy. Deep learning and Random Forest models are employed to identify NFC tags, while uncertainty quantification, particularly Conformal Prediction, accelerates the identification process with high confidence and precision. A software-defined radio (SDR) testbed is developed to transmit customized commands and collect multi-channel multi-rate NFC signals. The multi-channel multi-rate NFC signals are progressively collected to ensure fast and accurate identification. Experimental results demonstrate that the proposed system achieves high accuracy by adaptively utilizing the optimal combination of NFC signals. The developed solution is model-agnostic which can be utilized for any machine learning-based NFC tag identification.   
\end{abstract}


\section{Introduction}
Near Field Communication (NFC) technology has become a vital component of our daily life which provides convenience and security across a wide range of applications. It is widely used in contactless payments, access control systems, and public transportation. The seamless data communication and secure transactions can be performed by putting two devices in close proximity. While NFC enhances user convenience, its widespread adoption has also raised security concerns, particularly due to the cloning attack which can clone NFC tags without authorization. As NFC technology becomes more ubiquitous, the threat of tag cloning and counterfeit tags compromising secure systems has intensified.

RF fingerprinting, which analyzes the unique physical layer characteristics of each tag's signal during communication, has been shown to be an effective approach to identify NFC tags \cite{8342911,electronics12030559,10229040}. Machine learning (ML) particularly deep learning (DL) models \cite{lee2021deep} have been extensively used to analyze RF fingerprinting data \cite{rajendran2022rf}. Although RF fingerprinting has been studied for various Internet of Things (IoT) device identification, including NFC tags and devices, existing NFC RF fingerprinting research mainly focuses on the ISO14443 tags (used for financial transactions with short ranges and high data rates i.e., higher than 106 kbps) while another widely used NFC protocol ISO15693 (used for access control with short to mid-range and low data rate, e.g., 6.62 kbps) has not been studied \cite{10229040,iso15a}. The communication protocols and parameters are different between ISO14443 and ISO15693. Since cloning attacks on ISO15693-compliant tags/cards can threaten the associated security applications, it is essential to study the RF fingerprinting of ISO15693-compliant tags/cards by leveraging the unique characteristics of the communication protocol. To study the RF fingerprinting of ISO15693, we need to address the following research challenges.

First, there is a lack of reconfigurable communication systems that can interrogate NFC devices to generate arbitrary responses. Although ISO15693 protocol defines multiple channels and multiple data rates, commercial readers only use one of those, which limits our capability to gather multi-channel multi-rate data for RF fingerprinting to increase the identification accuracy. It is imperative to develop a reconfigurable system to collect arbitrary responses. Also, the high volume of data required for effective RF fingerprinting demands high storage and processing capacity, which adds computational and logistical burdens. Addressing these challenges necessitates standardized data collection protocols and access to diverse NFC tags, which ensures that models trained on these datasets generalize effectively to real-world scenarios.
 
Second, the identification results using ML/DL models have limited interpretability and the certainty of the decision cannot be effectively quantified. The lack of interpretability limits the applicability of RF fingerprinting, as its decision-making process remains unknown, making it difficult to predict or explain misclassifications. The inability to extract meaningful insights into the key RF characteristics of tags further complicates model validation and real-world deployment. Without clear interpretability, the trust in RF fingerprinting for security-critical applications is hindered. In addition, since errors in security applications are costly, it is important to evaluate the ML/DL models' certainty on the output. Decisions with high uncertainty can be further processed by collecting more data or resorting to human intervention. 

Third, the multi-channel multi-rate RF fingerprinting incurs additional power consumption and latency due to data collection and it is important to obtain an optimal way to minimize the data collection cost. Although the multi-channel multi-rate data improves identification accuracy, it requires the NFC device to generate multiple different responses. In NFC protocols, each NFC device response can only use a single predefined channel and data rate.   

Last, the received NFC signal power is highly sensitive to distance change and metallic environments. NFC uses a magnetic-based communication mechanism and the signal power decreases fast as the distance increases \cite{guo2021internet}. Also, any metals in the proximity of NFC readers or tags can detune their antennas, which incurs efficiency loss. These variations can alter the power of the received signal, leading to inconsistencies in the collected data. Consequently, these dynamic setup conditions may affect the accuracy and reliability of NFC tag identification, posing challenges for robust RF fingerprinting in real-world scenarios. 

NFC uses 13.56 MHz ISM band and the signal propagation is different from UHF IoT applications, e.g., WiFi, LoRaWAN, etc. As a result, some research challenges for UHF IoT applications do not exist in NFC. For example, NFC has a relatively short communication range and long wavelength, it does not experience significant multipath fading. 
 \begin{figure}[t]
     \centering
     \includegraphics[width=\linewidth]{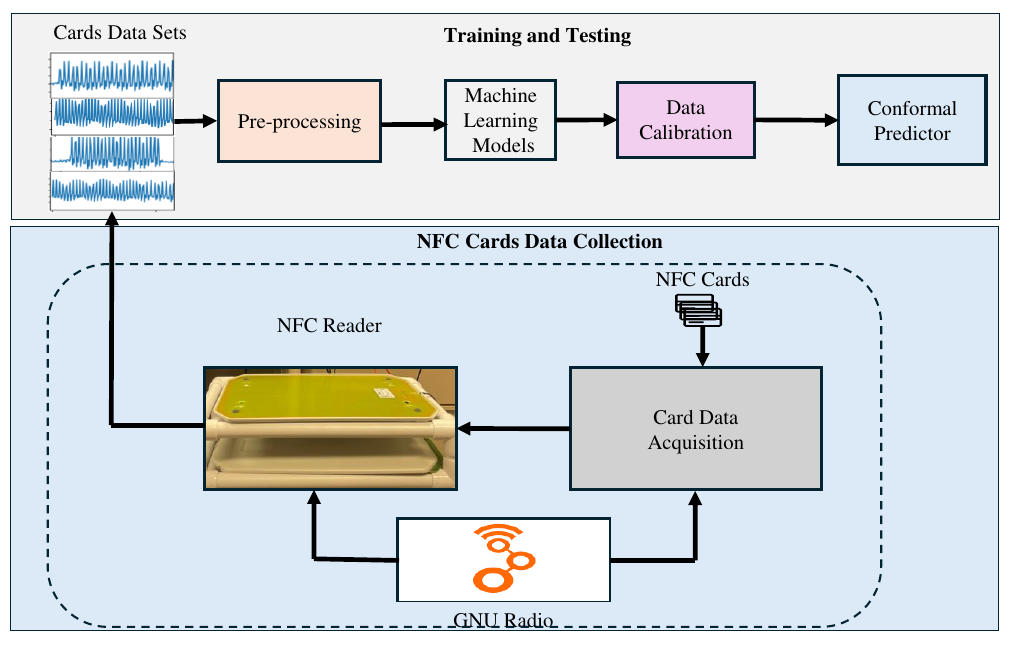}
     \caption{Workflow of the proposed RF fingerprint-based identification system.}
     \label{fig:workflow}
 \end{figure}

In this paper, we develop a multi-channel multi-rate NFC tag identification system with uncertainty quantification to address the aforementioned challenges. In particular, we leverage the multi-channel multi-rate modulation protocol defined in ISO15693 to send different interrogate commands to NFC tags and obtain 4 different types of responses. We develop a reconfigurable testbed using software-defined radios and GNU Radio to gather response data. The reader system is used to create a high-quality and well-structured dataset from ISO15693-compliant tags. This dataset is specifically tailored for integration into advanced ML/DL models, enabling the system to automatically extract discriminative features and make precise classification decisions. To ensure the creation of a robust and optimal dataset, we design an automatic data acquisition framework.

Then, we train a Random Forest model and a Convolutional Neural Network with ResNet and LSTM architecture (Advanced CNN model) using the multi-channel multi-rate NFC tag responses with various combinations. After that, we incorporate model-agnostic Conformal Prediction-based uncertainty quantification to evaluate the certainty of each identification result as shown in Fig. \ref{fig:workflow}. Although we can gather 4 types of NFC tag responses, not all of them are needed to identify a tag in practice. We design an adaptive interrogating system where the NFC reader uses uncertainty quantification results as an indicator to adaptively gather more tag responses. If the certainty is sufficiently high, the reader terminates the interrogation process and makes an identification decision using the data that has been collected. This ensures a fast, accurate decision. Since the uncertainty quantification is model-agnostic, the developed approach is generic, which can be used for any ML/DL based multi-channel multi-rate RF fingerprinting. The experiments validate the efficacy of the proposed solution using 52 ISO15693-compliant cards.    

The major novelty and contributions of this paper include the following. First, we develop a multi-channel multi-rate NFC reader system following ISO15693 protocol using software-defined radios. The system is fully reconfigurable and the software code and hardware configurations are shared in \cite{ISO15693NFCFingerprint}. Second, we develop ML/DL models with uncertainty quantification for NFC tag identification. The ML/DL models can output the identification results together with certainty estimation. This allows the user to understand the confidence level of the output, which is essential for security applications. Last, we design an adaptive NFC tag identification system using minimal NFC tag responses with high accuracy. The system adaptively collects data from the NFC tag without exhausting all combinations of channels and data rates.   

The remainder of this paper is organized as follows: Section II reviews existing device identification scenarios and related research. Section III introduces the ISO15693 protocol and its implementation using software-defined radios. Section IV presents the ML/DL models, along with uncertainty quantification using Conformal Prediction. Section V discusses the experimental results and findings. Finally, Section VI concludes the paper.

\section{Related Work}

\begin{table*}[t]
        \centering
        \captionsetup{justification=centering, font=small, skip=0pt}
        \caption{\scshape \\ Comparison With Previous Works}
        \begin{tabular}{c c c c c}
            \specialrule{.1em}{0em}{0em} 
            \textbf{Work} & \textbf{Protocol} & \textbf{Scale
} & \textbf{Model} & \textbf{Metric} \\
            \hline       
  \cite{electronics12030559} & ISO 14443-A & 4 & CNN  &  F1-score of 98.89\%, Average accuracy of 98.89\%\\
           \cite{10229040}  & ISO 14443 &  600 & Contrastive NN & FRR=3.7\%, FAR=4.1\% \\
           \cite{lee2021deep}  & ISO 14443 &  50 & FNN, RNN, CNN & Accuracy=96.16\%, 96.09\%, 94.99\% \\
            \textbf{This work} & {ISO 15693 }
 & {52} & {Random Forest} &  Adaptive Identification Accuracy=85.81\% \\
 \textbf{This work} & {ISO 15693 }
 & {52} & {Advanced CNN} & Adaptive Identification Accuracy=95.97\% \\
            \specialrule{.1em}{0em}{0em} 
        \end{tabular}
        \label{table:compare_works}
    \end{table*}

Various fingerprint-based tag identification methods have been proposed on different platforms, as summarized in Table \ref{table:compare_works}. In \cite{electronics12030559}, ISO/IEC 14443-A is explored. It introduces a DL-based RF fingerprinting method for NFC relay attack detection. Using a CNN classifier, the approach analyzes a dataset containing normal, wired, and wireless relayed signals from four NFC smart tags. In \cite{10229040}, it introduces a protocol-agnostic NFC fingerprinting scheme that leverages multi-frequency tag responses to extract unique tag characteristics. They analyzed response amplitudes from 600 ISO 14443-compliant tags and utilized contrastive learning for authentication. Furthermore, the work in \cite{lee2021deep} implements an NFC authentication system to capture one-bit RF signals for feature extraction. Three DL models including fully connected neural networks (FNN), convolutional neural networks (CNNs), and recurrent neural networks (RNNs) are utilized to distinguish 50 NFC tags. A comparison of the aforementioned research and this paper is given in Table~\ref{table:compare_works}. In this paper, the NFC tags are ISO15693-compliant and the difference from ISO14443 is introduced in Section~\ref{sec:background}.  

Traditional RF fingerprinting methods rely on manually crafted features designed for specific device classes and protocols \cite{10.1145/1409944.1409959,5934926,4550352}. These approaches often involve computationally intensive feature extraction and selection, which limits their scalability and adaptability. In contrast, modern advancements leverage DL techniques \cite{article}, enabling automatic extraction of complex features without the need for application-specific customization \cite{8054694,8509635,8882379,8737459}. To demonstrate the model-agnostic nature of our fast identification system, we employ both Random Forest and a DL-based Advanced CNN architecture in this study.  

RF fingerprinting features can be extracted from In-phase and Quadrature (IQ) data samples. However, as noted in \cite{Riyaz2018DeepLC}, the lack of standardized datasets for training and evaluation remains a significant challenge, alongside the difficulty of determining optimal partition lengths for input time-series data. Additionally, the development of robust datasets and the selection of discriminative features to distinguish individual devices are critical challenges. Several studies have explored different feature extraction approaches. For instance, \cite{electronics11020269,8631016} propose energy criterion-based and transient duration-based methods to detect and extract features from the signal preamble during power-on. Similarly, \cite{article123} focuses on extracting statistical features from the signal preamble, aiming to reduce computational overhead while preserving classification accuracy. In \cite{268896}, a passive fingerprinting method is introduced for identifying wireless device drivers operating on IEEE 802.11-compliant nodes. This method involves capturing probe request frames from devices and applying a supervised Bayesian model to analyze traces and generate unique fingerprints for device drivers. 

For security applications, it is crucial for the ML/DL models to indicate their confidence or certainty on decisions. Recent studies have explored adaptive trust mechanisms and physical-layer security to enhance device authentication. The research in \cite{8955830} proposed an adaptive trust management scheme that integrates soft authentication and progressive authorization to dynamically evaluate the trustworthiness of transmitters. By leveraging online Conformal Prediction and ML algorithms, this system continuously adjusts trust levels based on time-varying physical layer attributes like carrier frequency offset (CFO) and received signal strength indicator (RSSI). This approach addresses key limitations of static, binary authentication models by ensuring continuous validation and improving system robustness in dynamic environments. 
The proposed solution in this paper also employs Conformal Prediction to achieve high-confidence identification of NFC tags in real-time. Conformal Prediction provides statistical guarantees about the reliability of predictions by estimating the certainty associated with classification decisions \cite{2024arXiv241111824A,10630545}. By leveraging Conformal Prediction in conjunction with traditional ML model (Random Forest) and advanced DL models (Advanced CNN model), our approach dynamically evaluates the likelihood of the data samples belonging to a specific tag.

\section{Background and Automatic Data Collection}
\label{sec:background}
In this section, we first explain the principles of multi-channel load modulation and its use as an NFC tag fingerprint. Next, we provide an overview of the ISO15693 NFC protocol, followed by a discussion on the automatic data collection process using customized software-defined radio programs.

\subsection{Creation of NFC Tag Fingerprints from Tag Responses}
\begin{figure}[t]
    \centering
    \includegraphics[width=0.49\textwidth]{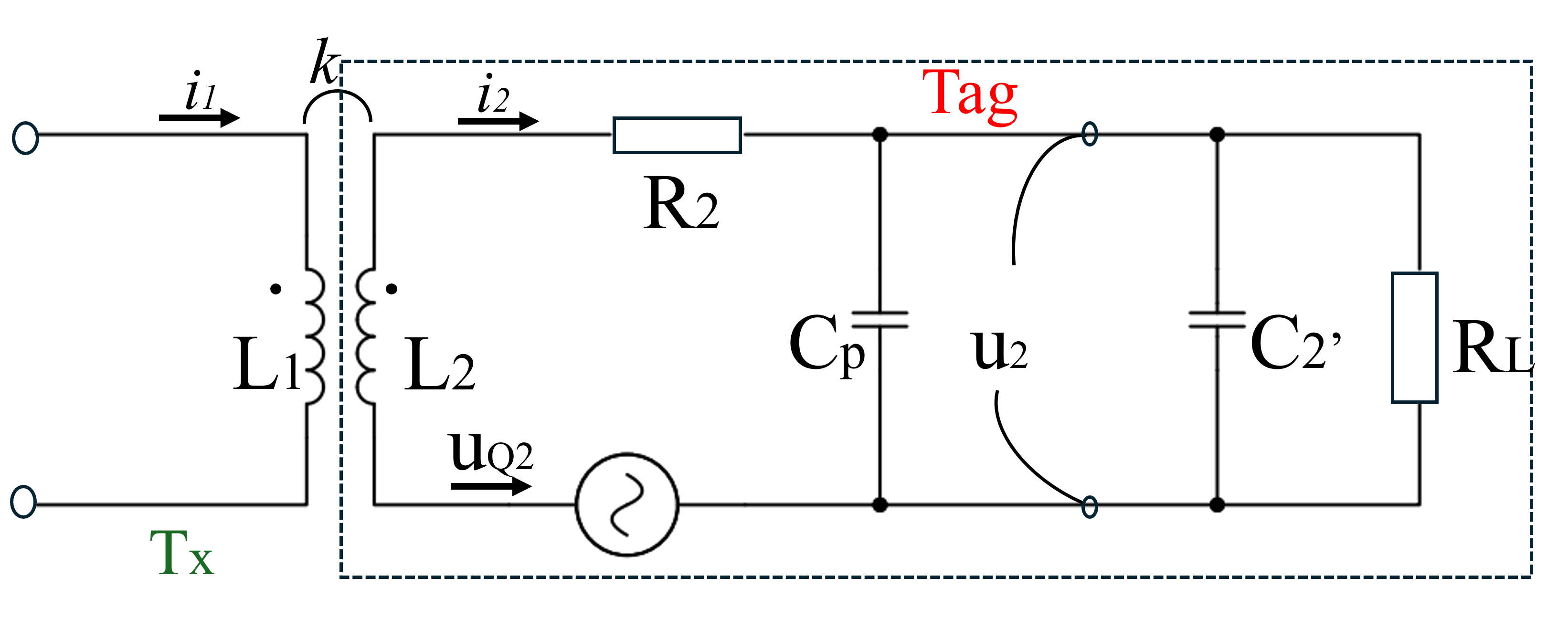}
    \caption{Equivalent circuit diagram for a reader and a tag.}
    \label{fig:tag_reader_equiv}
\end{figure}

NFC uses close coupling systems which are designed to operate within a range of 1 cm to a maximum of about 10 cm. As shown in Fig.~\ref{fig:tag_reader_equiv}, the reader (TX) and tag antennas ($L_1$ and $L_2$) are closely coupled through magnetic coupling. The reader sends out interrogate signals which induce a current $i_2$ and a voltage $u_2$ in the tag. Tags are classified as active or passive based on their power supply. Active tags have an internal battery, using voltage $u_2$ primarily as a wake-up signal to activate the data carrier when it exceeds a threshold, after which the device enters operational mode. It returns to standby or sleep mode once the communication is complete or if $u_2$ falls below the minimum level. In contrast, passive tags derive all their power from $u_2$, which is converted into direct current using a low-loss bridge rectifier and smoothed to operate the data carrier. The NFC reader initially transmits a high-power continuous wave (CW) signal to activate the NFC tag. Then it sends a request data packet. In response, the NFC tag harvests energy from the CW signal and subsequently transmits its response. The NFC tags used in this paper are in passive mode.

The harvested voltage $u_2$ by the passive tag is expressed below as \cite{10.5555/861917}:
\begin{equation}
u_2 = \frac{\omega \cdot k \cdot \sqrt{L_1 L_2} \cdot i_1}
{\sqrt{\left(\frac{\omega L_2}{R_L} + \omega R_2 C_2\right)^2 + \left(1 - \omega^2 L_2 C_2 + \frac{R_2}{R_L}\right)^2}}
\end{equation}
where $\omega=2\pi f$, $f$ is the carrier signal frequency, $C_2=C_2'+C_p$, $L_1$ and $L_2$ denote the self-inductance of the reader's and tag's coil antennas, respectively, $i_1$ is the current of the TX antenna, and $R_2$ and $R_L$ are the coil resistance of the tag and the load resistance, respectively.

 \begin{table*}[t]
    \centering
    \captionsetup{justification=centering, font=small, skip=0pt}
    \caption{\scshape \\ Response Modes With Subcarrier And Data Rates Settings \cite{iso15a}} 
    \begin{tabular}{c c c c c c c }
        \specialrule{.1em}{0em}{0em} 
        \textbf{Mode} & \textbf{Command} & \textbf{Data Size} & \textbf{Data Rate} & \textbf{Subcarriers} & \textbf{Subcarrier One Frequency} & \textbf{Subcarrier Two Frequency}\\
        \hline
        OneHigh & 0x2601000AF6 & 910 samples & 26.48 kbits/s  & Single & 423.75 kHz & --\\
        OneLow & 0x240100BF4E & 3624 samples & 6.62 kbits/s  & Single &  423.75 kHz & --\\
        TwoHigh & 0x270100502A & 900 samples & 26.69 kbits/s  & Dual &  423.75 kHz & 484.28 kHz\\
        TwoLow & 0x250100E592 & 3599 samples & 6.67 kbits/s & Dual &  423.75 kHz & 484.28 kHz\\
        \specialrule{.1em}{0em}{0em} 
    \end{tabular}
    \label{tab:examp2}
    \vspace{1pt}
\end{table*}

The harvested voltage $u_2$ measures the amount of energy harvested by the tag, which can be reflected from the superposition of reader CW and the tag response signals as expressed in the form \cite{10229040}:
\begin{equation}
    \begin{aligned}
        S_{tag}(f) = A_{CW}(f)  \delta(\pm f) + A_{tag}(f) \delta(f \pm f_s), 
    \end{aligned}
\end{equation}
where $A_{cw}(f)$ denotes the amplitude of CW signal, $A_{tag}(f)$ represents the amplitude of the modulated tag's responses, $f_s$ denotes sub-carrier frequency from tag's load modulation.
The Dirac delta function is:
\begin{align}
\delta(\pm f) = \frac{e^{j 2 \pi f t} + e^{-j 2 \pi f t}}{2},
\end{align}
which represents the impulse response at the CW frequency ($f$=13.56 MHz) and
$\delta(f \pm f_s)$ represents the tag load modulation spectral components at $f \pm f_s$, where $f_s=f/32$ or $f/28$.

The multi-channel multi-rate approach provides unique fingerprints for NFC tags due to two key properties. First, hardware variations arise from manufacturing imperfections. Even when produced by the same manufacturer, NFC tags exhibit distinct physical characteristics due to inherent inconsistencies in the manufacturing process. These imperfections result in variations in $L_2$, $C_2$, $R_2$, and $R_L$, which in turn cause $u_2$ and $s_{tags}(f,t)$ to differ uniquely across tags. Second, NFC tags exhibit distinct signal responses at different subcarrier frequencies and communication data rates as specified in ISO15693 \cite{iso15a}. By capturing and analyzing multi-channel multi-rate signals, we can uniquely identify NFC tags based on these variations.

In real-world scenarios, the received tag response signals are influenced by the frequency response $H(f)$ of the reader's RF front-end, which includes both the RF circuitry and the antenna. Meanwhile, the NFC tag is powered through inductive coupling, where the amount of harvested energy depends on the coupling coefficient ($k$) between the reader antenna and the tag coil \cite{10.5555/861917}. Thus, the actual received tag response signal is expressed as:
\begin{equation}
         R_{tag}(f) = H(f) \cdot k \cdot (A_{CW}(f)  \delta(\pm f) + A_{tag}(f) \delta(f \pm f_s)) 
\end{equation}

Data sent from the NFC tag to a reader is achieved through load modulation with a modulated subcarrier signal. This process involves switching the resistive or capacitive modulation element on and off in synchronization with the subcarrier frequency. The subcarrier signal is further modulated in time using a Manchester-coded data stream with Amplitude Shift Keying (ASK) or Frequency Shift Keying (FSK) modulation \cite{10.5555/861917}. The reader controls the modulation scheme by setting a flag bit in the transmission protocol header, as specified in the protocol \cite{iso15a}, requiring the NFC tag to support both modulation methods. Additionally, the data rate is adjusted between two predefined values, with the reader selecting the rate through another flag bit in the protocol header. Modulation of these subcarriers along with multi-rate settings can gather comprehensive responses of NFC tag to create unique fingerprints. 

Note that, the impact of noise from RF devices and received signal strength or Signal-to-Noise (SNR) ratio can also affect the performance. The signals acquired by generic RF devices are inherently affected by random noise, including thermal and flicker noise. These noise sources introduce instantaneous variations in repeated measurements of $s_{tag}(f,t)$, even for the same NFC tag. Such variations may influence the distribution of extracted fingerprints, causing the clusters of different tags to become less distinct and more closely spaced. As will be discussed later, existing NFC only works in a short range and the interference and multipath fading are negligible which leaves the SNR as the dominant factor affecting collected signal quality.


\subsection{NFC Transmission Protocol}

The NFC protocol utilized in this research is the ISO15693, which is designed for communication with vicinity integrated circuit cards (VICCs) and vicinity coupling devices (VCDs) \cite{iso15a,Ansari2024NFC}. In this paper, we refer to the VICC as an NFC tag and the VCD as a reader. In the ISO15693 protocol, the reader generates an RF operating field to power the NFC tags. This operating field, which has a carrier frequency of 13.56 MHz $\pm$ 7 kHz, provides the energy required for the passive NFC tags to function. 

The modulation scheme used is ASK, and the depth of ASK modulation can vary between 10\% and 100\%, depending on the system requirements. For our data collection process, we use 100\% ASK modulation. There are two pulse position modulation schemes used for reader-to-tag communication. For data encoding in this paper, we adopted the 1-out-of-4 coding scheme as specified in the ISO15693 standard. In this scheme, the relative position of a pulse represents two bits of information. This efficient encoding method allows for reliable communication with minimal errors.

When the NFC tag transmits data back to the reader, it also employs ASK modulation. The tag can respond using either one or two subcarriers, depending on the mode of communication. The primary subcarrier operates at a frequency of 423.75 kHz, while the secondary subcarrier, when used, operates at 484.28 kHz. Also, the protocol defines two data rates, i.e., 26.48 kbps and 6.62 kbps using different signals. Usually, the NFC tag identification only adopts one type of signal. In this paper, we consider all combinations of subcarriers and data rates. Table~\ref{tab:examp2} presents the response modes, the corresponding hexadecimal commands, subcarrier and data rate settings, as well as the number of samples captured for each of the four primary tag response modes. We define the response modes based on the subcarrier and data rate settings. The ``One'' indicates a single subcarrier (423.75 kHz), while ``Two'' represents dual subcarriers (423.75 kHz and 484.28 kHz). ``High'' denotes a high data rate (~26.48 kbit/s, ~26.69 kbit/s), whereas ``Low'' refers to a low data rate (~6.62 kbit/s, ~6.67 kbit/s). Thus, 
\begin{itemize}
    \item {\bf OneHigh} represents one subcarrier and high data rate.
    \item {\bf OneLow} represents one subcarrier and low data rate
    \item {\bf TwoHigh} represents two subcarriers and high data rate.
    \item {\bf TwoLow} represents two subcarriers and low data rate.
\end{itemize}

\subsection{Automatic Data Collection using Software-Defined Radios}

\begin{figure}[t]
    \centering
    \includegraphics[width=0.45\textwidth]{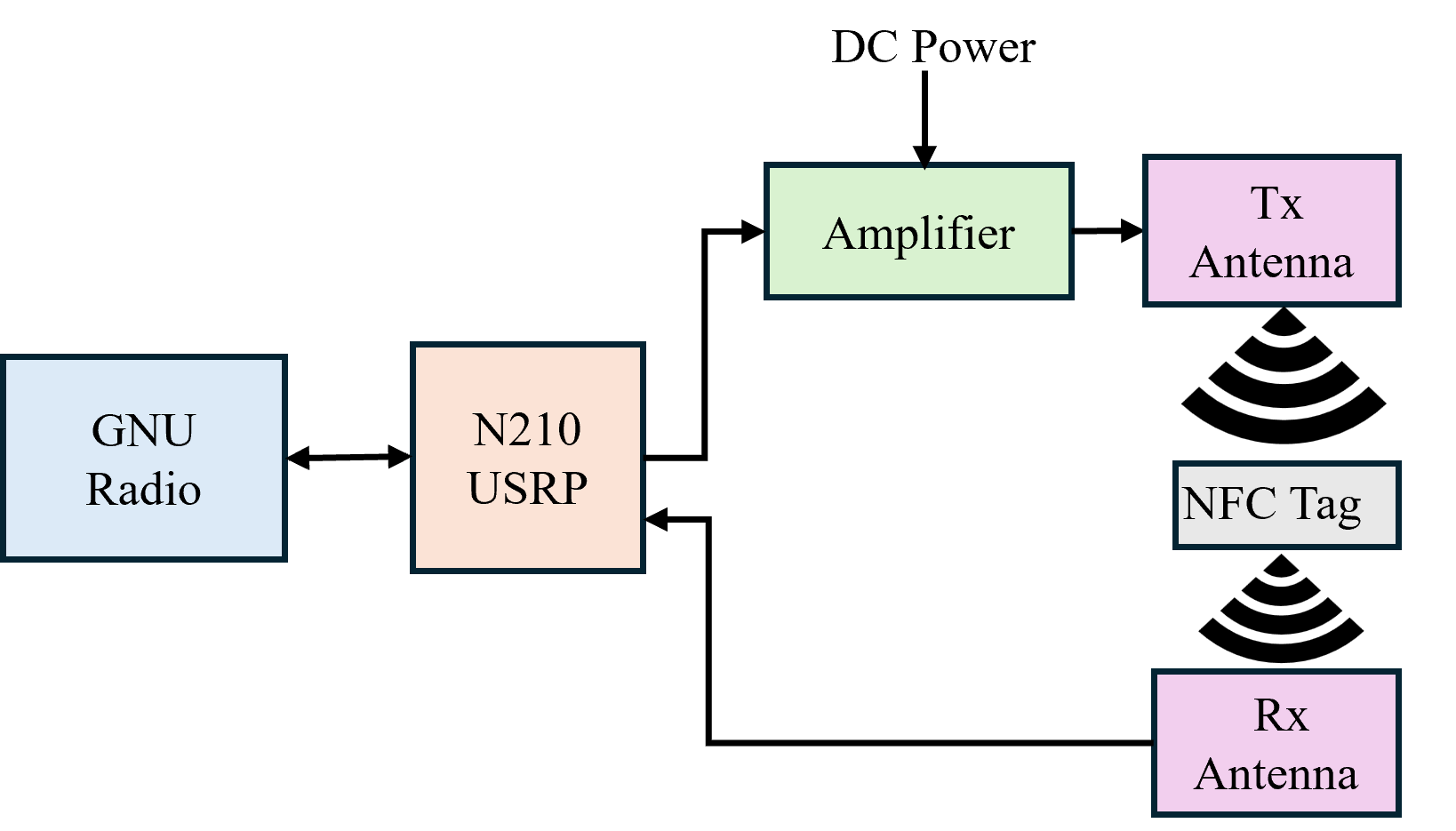}
    \caption{Experiment setup block diagram.}
    \label{fig:setup_diagram}
    \vspace{-5pt}
\end{figure}

\begin{figure}[t]
    \centering
    \includegraphics[width=0.45\textwidth,height=6.4cm]{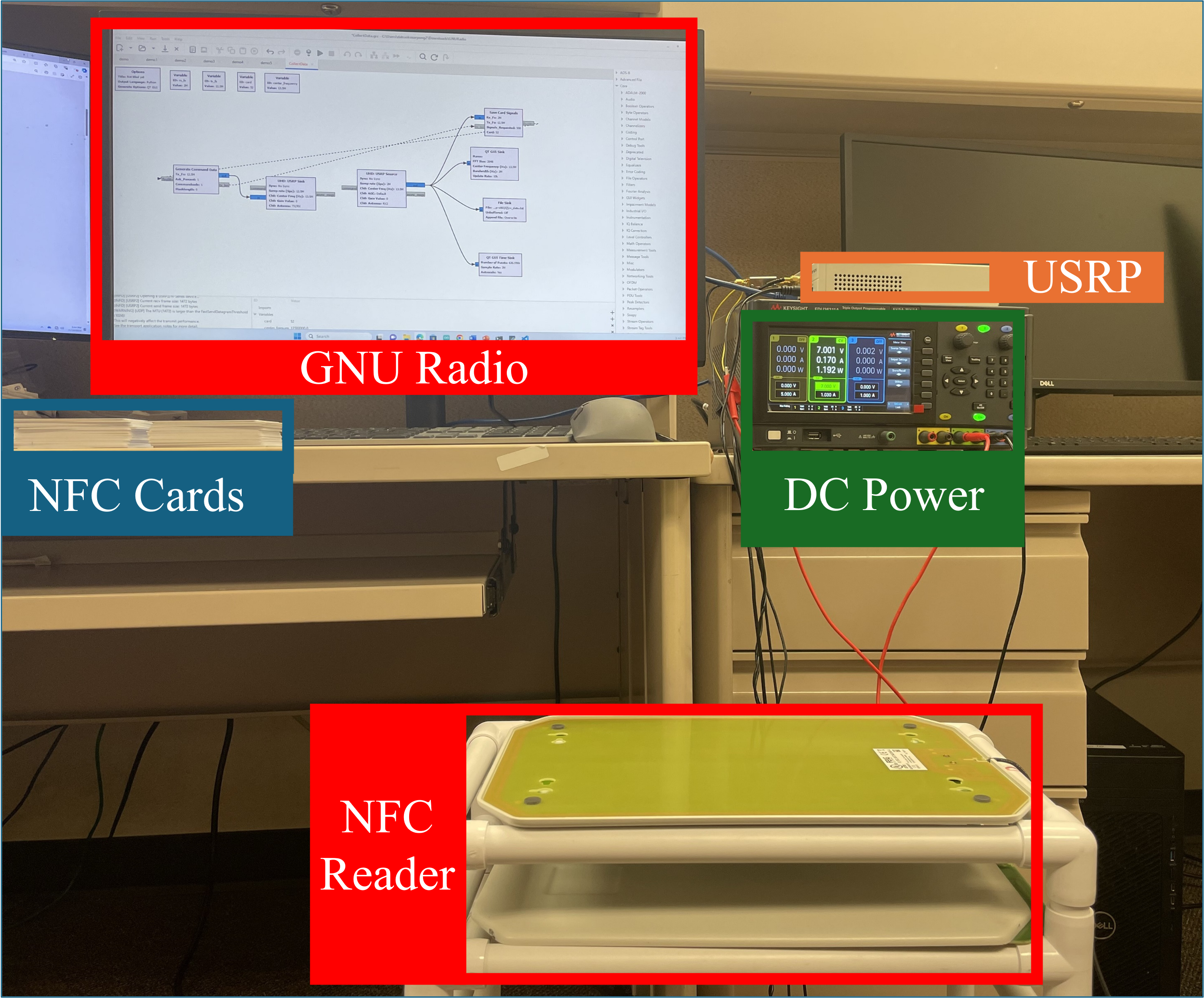}
    \caption{Lab environment experiment setup.}
    \label{fig:setup_image}
    \vspace{-5pt}
\end{figure}

 Existing ISO15693 readers usually only allow one of the four modulation types. To receive all combinations of tag signals, we develop a reconfigurable NFC communication system utilizing software-defined radios. The Ettus N210 USRPs (Universal Software Radio Peripheral) are used as the main hardware platform for transmitting data to and receiving responses from NFC tags. Our setup includes 52 ISO15693 NFC tags, from which we collect responses. Two FEIG ISC.ANT340/240-A antennas are connected to the N210 radios, with one dedicated to transmission and the other to reception. An illustration of the system is shown in Fig.~\ref{fig:setup_diagram} and the associated implementation is given in Fig.~\ref{fig:setup_image}. 

To enhance the signal strength of the USRP transmission, we incorporated a Mini-Circuits ZX60-100VHX+ 12V amplifier between the USRP and the transmission antenna. This amplifier is essential, as the transmission power of the USRP is insufficient to activate NFC tags without additional amplification. The amplified signal ensures reliable tag activation and response, facilitating effective data collection for our experiments.
    
GNU Radio serves as the software interface for controlling the Ettus N210 USRP, enabling the transmission and reception of data in our NFC system. It facilitates the transmission of ISO15693 command data from the USRP to the NFC tag and simultaneously receives responses from the tags. The transmission sampling rate is configured at 12.5M samples/second, while the reception sampling rate is set at 2M samples/second. 

According to the ISO15693 standard, each NFC tag can respond in 4 distinct modes, determined by combinations of subcarrier settings (one or two subcarriers) and data rates (high or low). The response mode of the NFC tags is configured through the flag bits included in the command requests sent from the reader to tags. Each response mode is associated with a unique hexadecimal command sequence and produces a different number of samples in the captured response. According to the ISO15693 protocol \cite{iso15a}, an inventory command request transmits between 40 and 112 bits to the NFC tag, depending on the mask length and the optional AFI field. In our setup, we used a mask length of zero and do not include the AFI, resulting in 40 bits being transmitted for each inventory request. The format of an inventory request is shown in Table~\ref{table:inventory_request}. Table~\ref{table:inventory_request} also provides an example of how the first command is constructed.

The key component of an inventory request is the flag field, which indicates how the NFC tag should respond to the command. The first two bits of the flag field are particularly important. Bit 0 determines whether the tag responds with one or two subcarriers: '0' indicates one subcarrier, and '1' indicates two subcarriers. Bit 1 controls the data rate, where `0` indicates a low data rate and `1` indicates a high data rate.

\begin{figure*}[t]
    \centering
    \begin{subfigure}[b]{0.48\textwidth}
        \centering
        \includegraphics[width=0.88\textwidth]{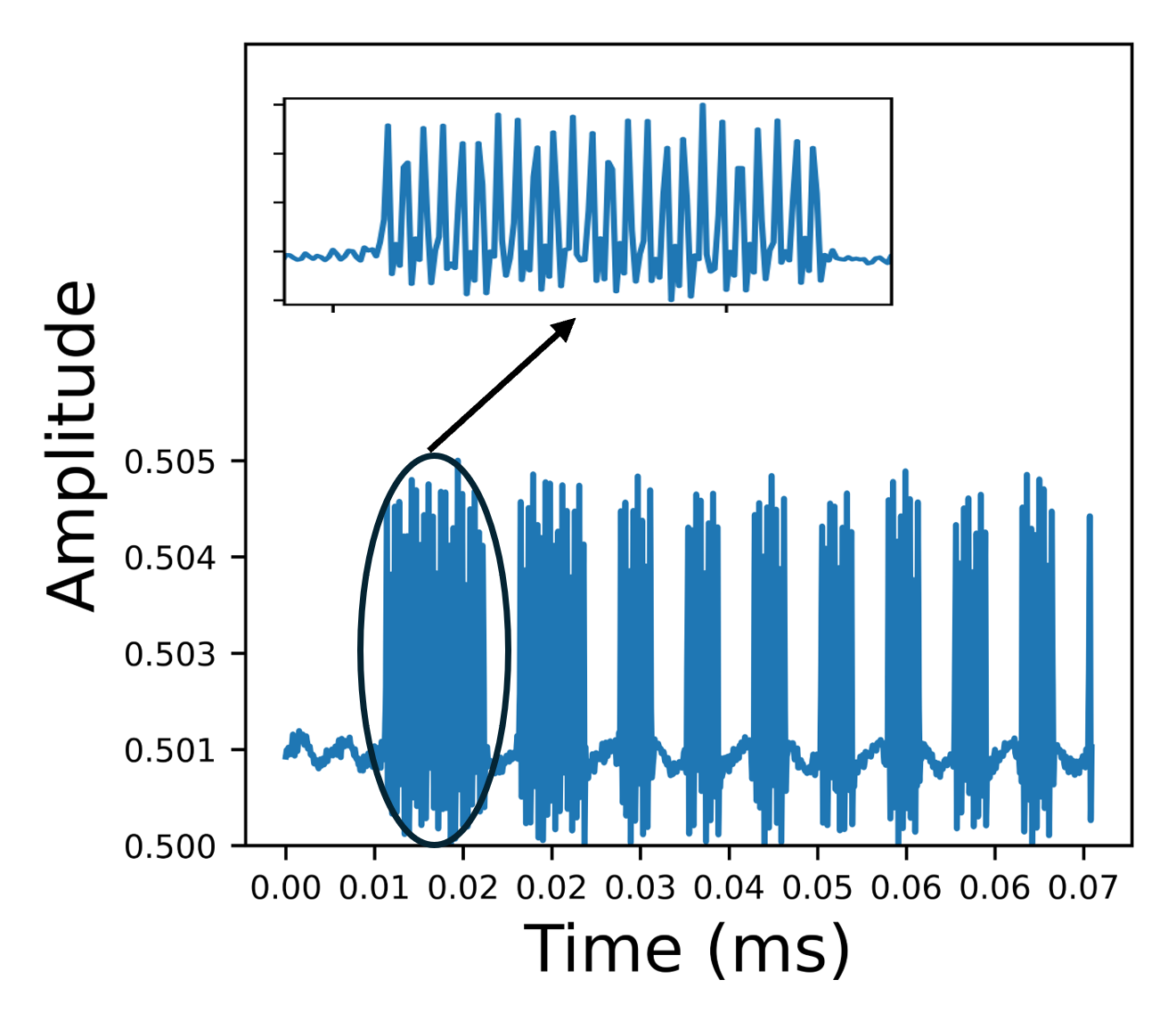}  
        \caption{One subcarrier high data rate.}
    \end{subfigure}
    \hfill
    \begin{subfigure}[b]{0.48\textwidth}
        \centering
        \includegraphics[width=0.88\textwidth]{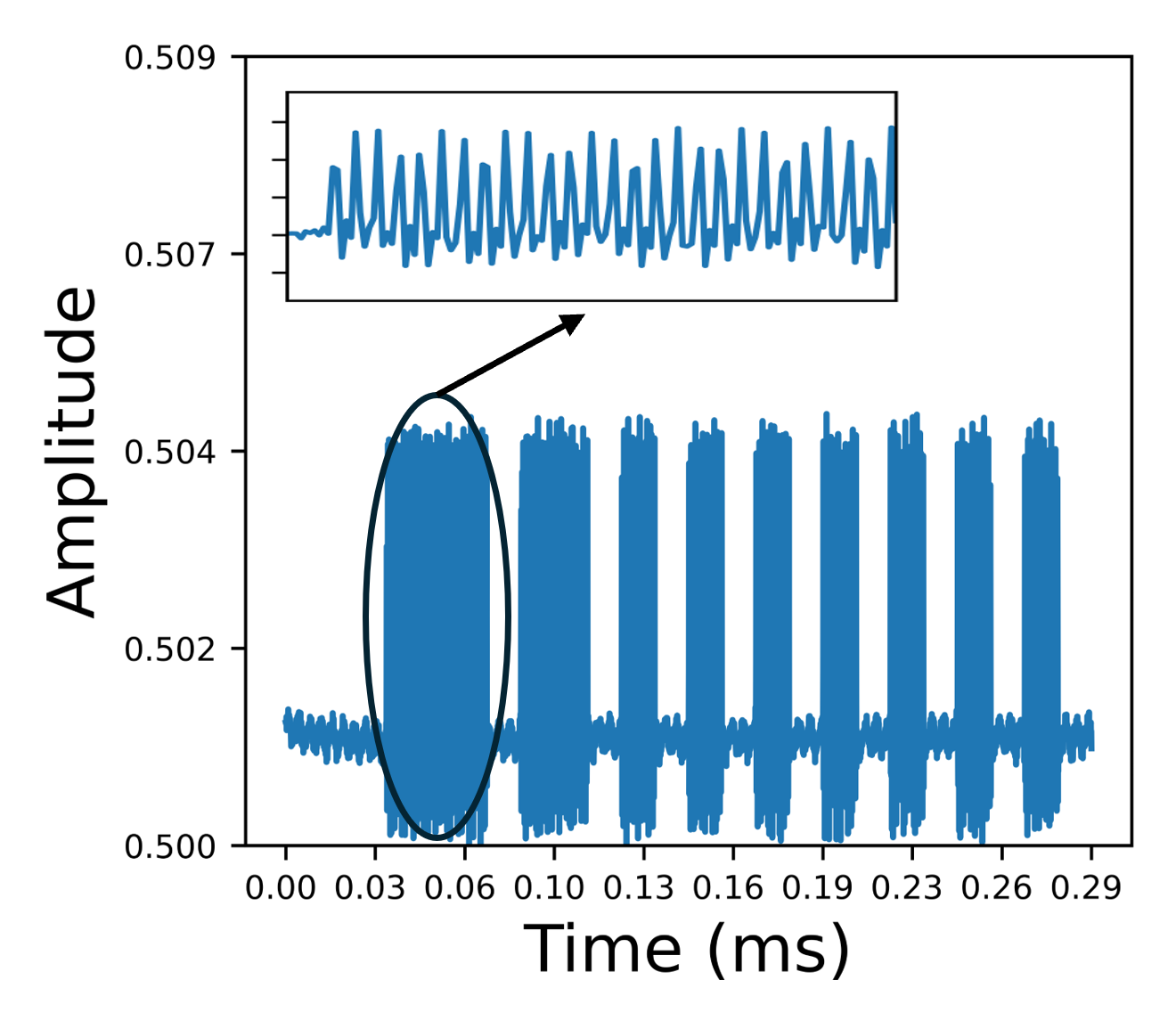} 
        \caption{One subcarrier low data rate.}
    \end{subfigure}
    
    \vspace{0.3cm} 
    
    \begin{subfigure}[b]{0.48\textwidth}
        \centering
        \includegraphics[width=0.88\textwidth]{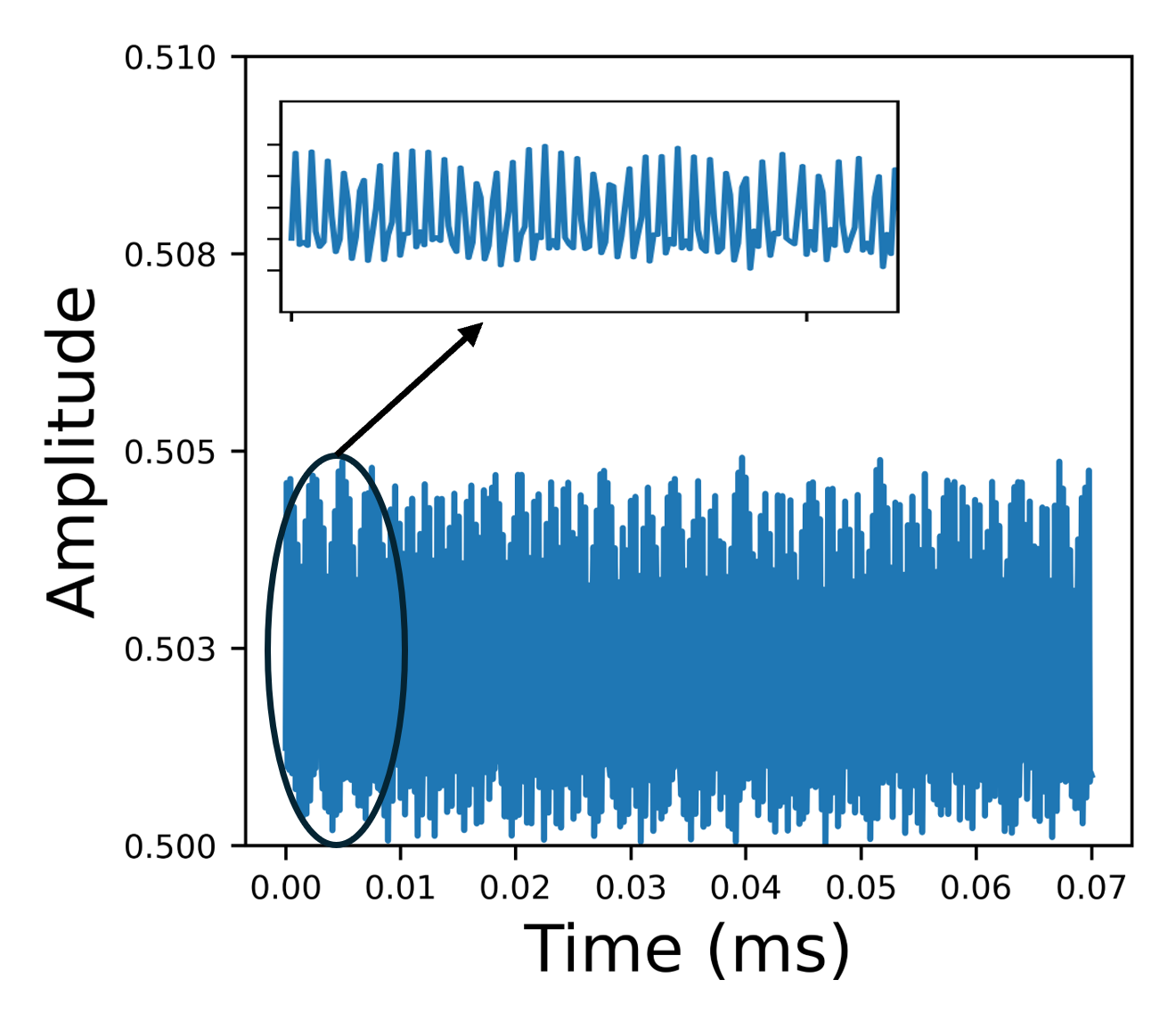}  
        \caption{Two subcarriers high data rate.}
    \end{subfigure}
    \hfill
    \begin{subfigure}[b]{0.48\textwidth}
        \centering
        \includegraphics[width=0.88\textwidth]{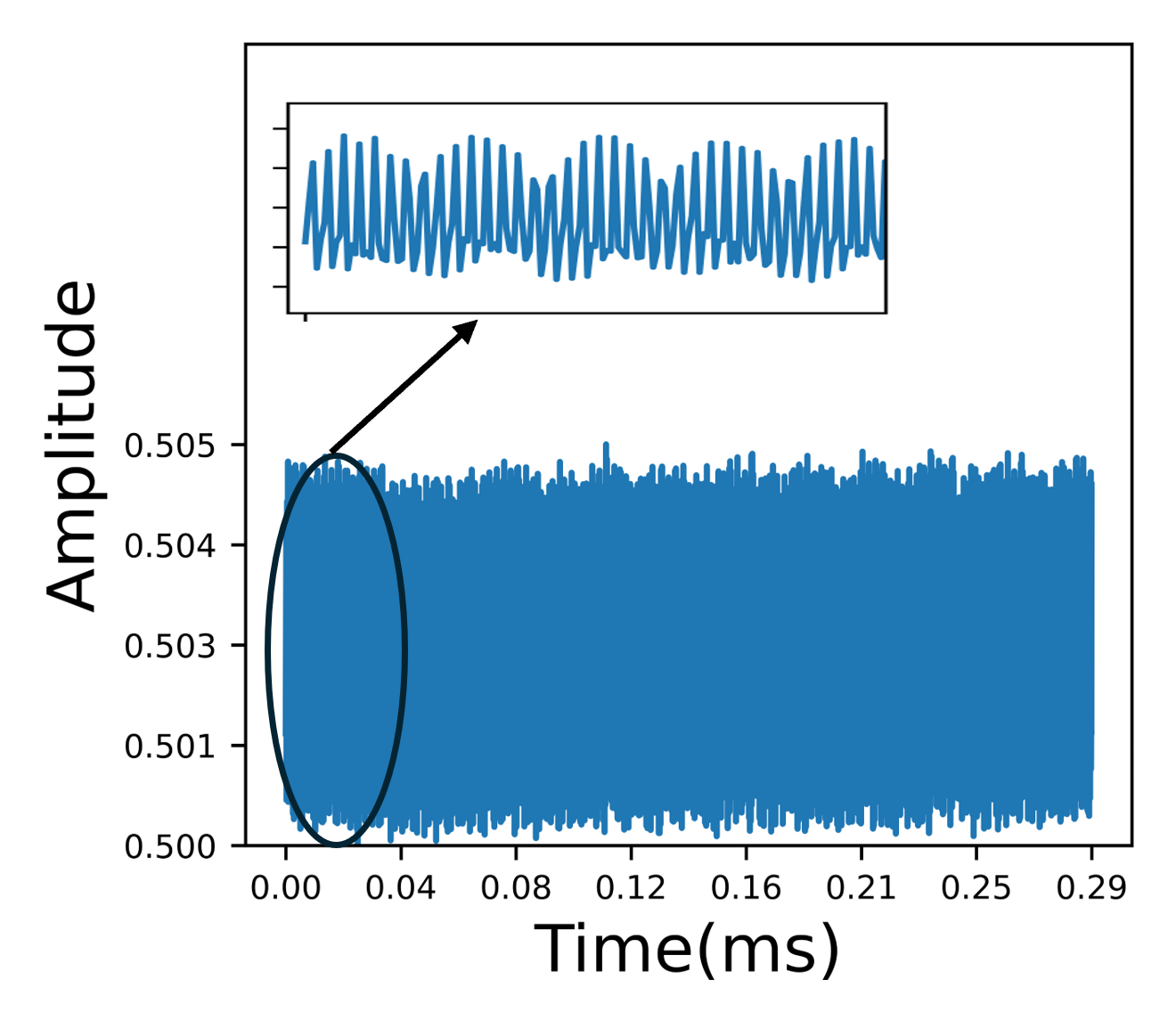} 
        \caption{Two subcarriers low data rate.}
    \end{subfigure}
    
    
    \caption{Responses for four ISO15693 response modes.}
    \label{responses_types}
\end{figure*}

As seen in Table \ref{table:inventory_request}, two primary elements change between each command: the flag bits and the CRC (Cyclic Redundancy Check) values. The flag bits are modified to collect data across the 4 different response modes. The CRC, represented by the first 4 hexadecimal values, adjusts accordingly each time the flag bits change, as the CRC needs to be recalculated to reflect the updated command. Figure~\ref{responses_types} provides a visual representation of the tag responses for each of these modes, highlighting the variations in their signal characteristics.

\begin{table}[t]
    \centering
    \captionsetup{justification=centering, font=small, skip=0pt}
    \caption{\scshape \\ Inventory Request Format and Example}
    \begin{tabular}{c c c}
        \specialrule{.1em}{0em}{0em} 
        \textbf{Field} & \textbf{Bits} & \textbf{Example} \\
        \hline
        Start of Frame & --- & --- \\
        Flags & 8 bits & 0x26 \\
        Command Code & 8 bits & 0x01 \\
        Optional AFI & 0 bits &  ---\\
        Mask Length & 8 bits & 0x00 \\
        CRC & 16 bits & 0x0AF6 \\
        End of Frame & --- & --- \\
        \specialrule{.1em}{0em}{0em} 
    \end{tabular}
    \label{table:inventory_request}
    \vspace{-15pt}
\end{table}
Upon receiving data, we develop a GNU Radio program to process the incoming signals to identify and extract the responses from the NFC tags, which are then saved for further analysis. We collected responses from 52 NFC tags, which are subsequently utilized to train ML and DL models, enhancing their ability to detect and distinguish between authorized and counterfeit tags based on subtle signal variations. Due to the limited communication range, the tag is placed on a designated surface above the receiving antenna for activation. By positioning the tag within the receiving antenna's range or directly on its surface, the tag's coil is precisely aligned within the air gap of a ring-shaped or U-shaped core. This configuration ensures efficient coupling, resembling the functional arrangement of a transformer between the tag and reader coils in Fig. \ref{fig:tag_reader_equiv}.

For each of the 52 NFC tags, we collected 2,000 responses, capturing 500 responses for each of the 4 response modes per tag. We focused on the first 8 bits of each response since these bits are consistent across all tags, indicating that the captured data is independent of the tag's UID. This approach resulted in a total of 104,000 data samples, providing a comprehensive dataset for evaluating the performance of our ML/DL models. The large volume of data enabled the models to effectively learn and distinguish the subtle variations in tag responses.

Before transmitting any command, we first pause sending any signals. This pause serves a critical purpose: it allows us to clearly identify when the transmission of the command begins, creating a distinct boundary in the signal. Once the RF field resumes and the command is transmitted, we can track the completion of the inventory request. Next, we employ a method to detect the start of the tag's transmission as it always lies just after the inventory request. This is crucial for accurately isolating the tag's response from the transmitted command. After identifying the start of the tag's response, we extract the desired portion of the signal—specifically, i.e., the first 8 bits of data. 

To ensure the accuracy and validity of the extracted response, we compare it to a known, pre-saved tag response that has the same number of samples. The comparison is done by calculating the Pearson correlation coefficient between the two responses. If the coefficient exceeds a predefined threshold, the response is considered valid. This validation step ensures that the extracted response is both reliable and consistent with expected tag behavior. Once validated, the response is saved to our dataset for further analysis.

\begin{figure}[t]
\centering
\includegraphics[width=0.8\columnwidth]{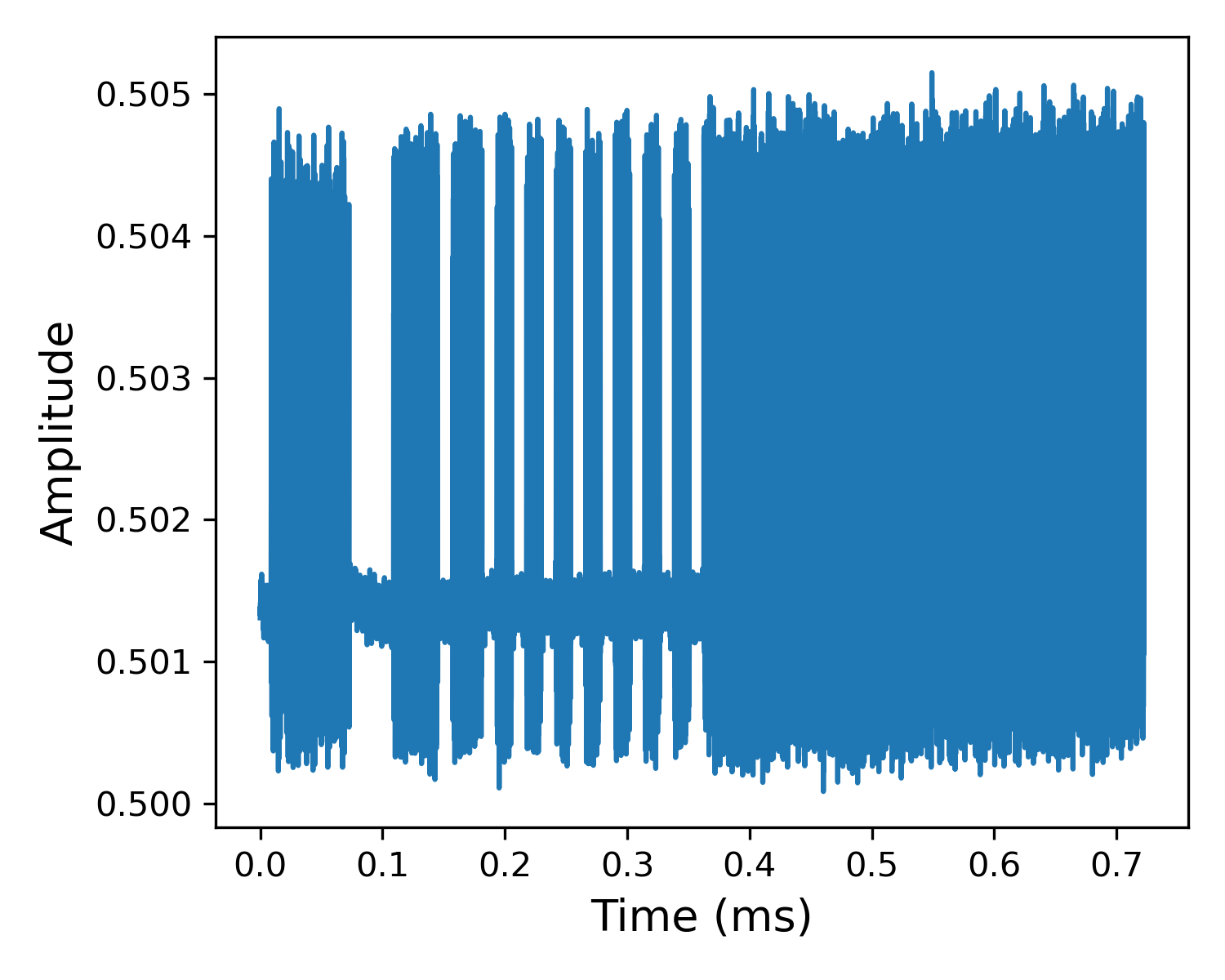}
\caption{Concatenated and normalized responses for a single NFC tag.}
\label{fig:concatenated_responses}
\vspace{-10pt}
\end{figure}
    
After data collection, the responses must be preprocessed before being used in ML/DL models. It was observed that different response modes exhibited varying amplitudes, and data collected on different days showed fluctuations in average signal strength. To mitigate these inconsistencies, normalization is applied to standardize the data. The normalization process involves selecting a reference point of 0.5. For each response, the difference between the smallest sample value and the reference point is computed. This difference is then subtracted from all samples, effectively shifting the amplitude of all responses to be centered around 0.5.

Once normalized, the responses from all modes are concatenated to form a comprehensive dataset, allowing the models to utilize all available data. The responses are organized into a 2-D matrix, with rows representing individual responses and columns representing the sample points of each response. A corresponding label array is also created to identify which rows belonged to each specific tag, facilitating accurate training and evaluation of the models. 

This paper aims to develop a fast, accurate solution, and thus we do not plan to use all combinations of the obtained modulated signals. We train and test models using a single type of response and their combinations. For datasets with multiple types of signals, we concatenate the signals in the time domain. An example of the concatenation of all 4 response modes for a single NFC tag is shown in Fig. \ref{fig:concatenated_responses}. The developed system will aggregate data samples by first collecting a single type of signal, e.g., OneLow, then progressively collecting more signal types until meeting the certainty requirements or exhausting all signal types. The datasets are prepared for training such ML/DL models.


\section{Conformal Prediction-Based Adaptive NFC Tag Identification}

In this section, we first implement a Random Forest model and a DL model which serve as baselines. The two models will be trained on each signal type and their combinations. Then, we introduce the Conformal Prediction model with adaptive data aggregation which can be used with any ML models. 

For the Random Forest and advanced CNN models, we train seven models using each approach. The $M_1$, $M_2$, $M_3$ and $M_4$ represent models trained using single channel and single data rate, including OneHigh, OneLow, TwoHigh and TwoLow signals, respectively. The $M_5$ (TwoTypes) model represent the model trained on the combination of OneHigh and OneLow datasets. Similarly, the $M_6$ (ThreeTypes) model is trained on combinations of OneHigh, OneLow and TwoHigh datasets. Finally, $M_7$ (FourTypes) is trained on the combination of OneHigh, OneLow, TwoHigh and TwoLow datasets.

\subsection{Random Forest Model}
A Random Forest Classifier is implemented using the RandomForestClassifier from the scikit-learn library in Python. To further enhance the robustness and reliability of predictions, we integrate MAPIE (Model Agnostic Prediction Interval Estimator) \cite{Cordier_Flexible_and_Systematic_2023}. MAPIE is a library designed to compute calibrated prediction intervals for ML models, providing statistical bounds on predictions. By wrapping the Random Forest Classifier with MAPIE, we are able to quantify the uncertainty associated with each prediction, ensuring more reliable and interpretable results. This approach not only improves the confidence in outputs but also aligns with the need for robust decision-making in a dynamic and uncertain environment.

To implement this, we first instantiate a random forest classifier using scikit-learn, followed by declaring a MAPIE classifier that wraps the Random Forest model. This provided us with both the classification model and the prediction intervals which will be used in Conformal Predictions. As mentioned earlier, we have four distinct response types from the NFC tags, and we divide the dataset into training and testing sets, using 70\% of the data for training and 30\% for testing and calibration (20\% testing and 10\% calibration).

To evaluate the performance of the Random Forest model, we trained seven different models to assess their accuracy. The first four models are trained using only one response type each: One Subcarrier High Data Rate, One Subcarrier Low Data Rate, Two Subcarriers High Data Rate, and Two Subcarriers Low Data Rate. In addition, we trained three more models that combined two (only one subcarrier signals), three (one subcarrier signals and TwoHigh signal), and all four response types, respectively, to see how the inclusion of more diverse data would affect model performance.

To measure the accuracy of each model, we used scikit-learn's `cross\_val\_score` function with a 3-fold cross-validation (cv=3). The average of these scores gives us an estimate of the model's accuracy across different subsets of the data, ensuring that the model's performance is not biased by the choice of a particular validation or test set.

\begin{figure}[t]
    \centering
    \rotatebox{270}{ 
        \includegraphics[width=0.49\linewidth]{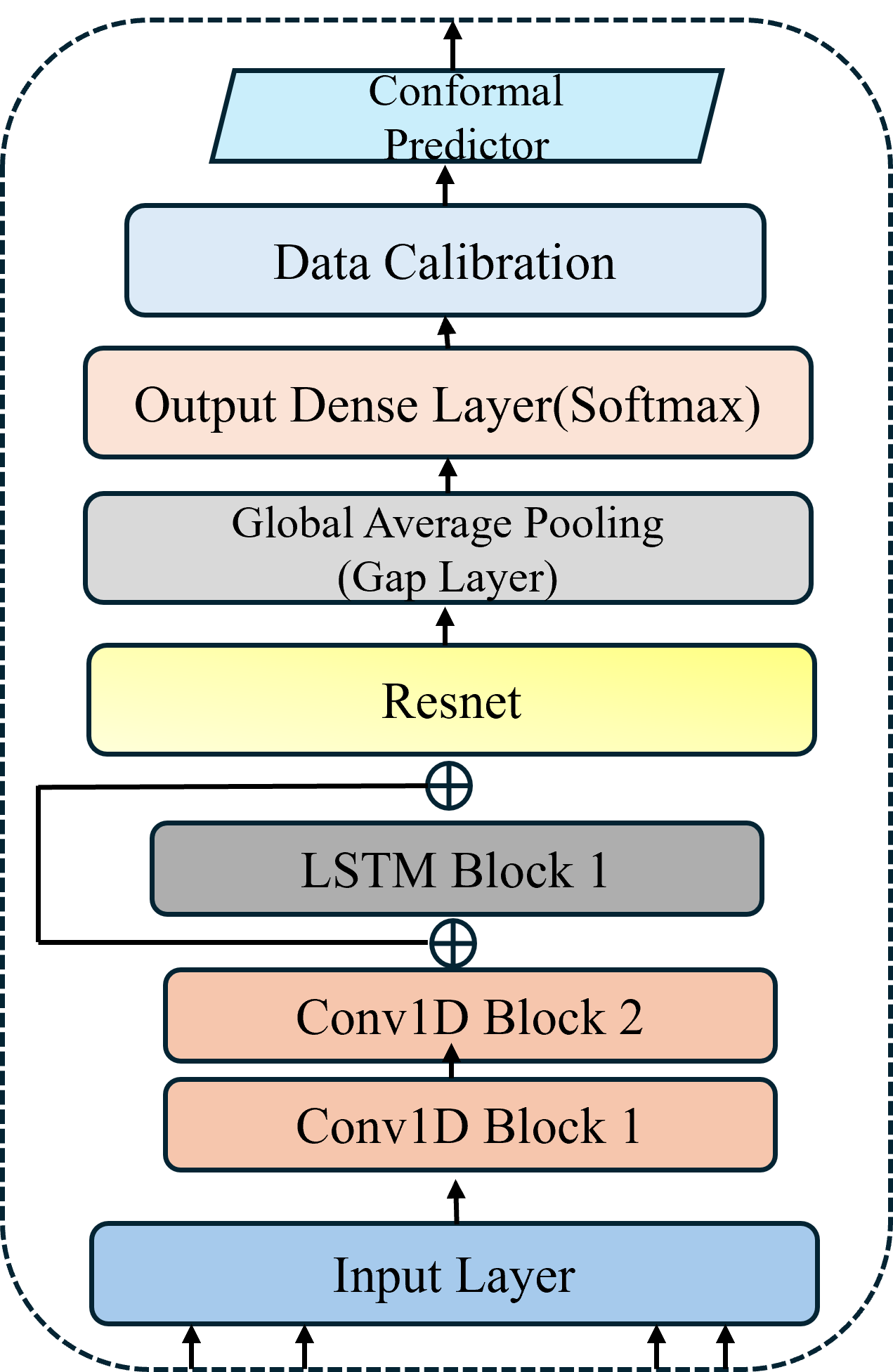}
    }
    \caption{Advanced CNN architecture for tag authentication.}
    \label{fig:CNNResnet_architecture}
\end{figure}

\subsection{DL Model with Advanced Convolutional Neural Network Architecture(Advanced CNN Model)}
Similarly, we employ a DL workflow for training a Convolutional Neural Network (CNN) with ResNet and LSTM architecture as shown in Fig. \ref{fig:CNNResnet_architecture} on NFC card data, followed by a Conformal Prediction process for logical implementation to assess model uncertainty. The signals are normalized using MinMaxScaler to scale their values between 0 and 1, ensuring consistency and improving model training. The structure begins with an input layer, which processes the data for subsequent transformations. The data flows through two Conv1D Blocks that extract local spatial features from the input using convolutional operations. Following these blocks, we incorporated an LSTM Block, which is adept at capturing temporal dependencies in sequential data, enhancing the model’s ability to understand patterns over time \cite{7508408}. A ResNet module is integrated to introduce residual connections, allowing the network to combat the vanishing gradient problem and enabling deeper network training by preserving information across layers. The extracted features are then passed through a Global Average Pooling (GAP) Layer, which reduces the feature map dimensionality and highlights the most prominent characteristics for classification. The Output Dense Layer (Softmax) produces probabilistic outputs across the target classes \cite{7820046}. 

To mitigate overfitting, a dropout rate of 50\% is applied to the fully connected dense layers and L2 kernel regularizer. The neural network is trained using the Adam optimizer with a learning rate of 0.0001. Additionally, a batch size of 32 is used to balance computational efficiency and model performance. The training process involves loading and normalizing NFC datasets, splitting data into 70\% training, 20\% testing, and 10\% calibration, and training the Advanced CNN model wrapped with KerasClassifier using MAPIE for the 7 dataset combinations. We train each model by setting the epoch number to 100 while applying an early stopping callback with a patience of 10 which stops training if validation accuracy does not improve for 10 epochs. Each model is trained over 3 trials, computing mean test accuracy and standard deviation, then saved for the Conformal Prediction phase.

\subsection{Conformal Prediction Modeling}
In this subsection, we employ the model-agnostic Conformal Prediction approach, which improves model reliability by generating an output label along with a prediction set that can be used to evaluate the certainty of the model's decision. A key advantage of this approach is its flexibility, as it can be integrated with any pre-trained model, making it model-agnostic. This ensures that the true tag identity is included within the predicted set ${\mathcal S}$ with a user-defined probability \cite{DBLP:journals/corr/abs-2107-07511}. We use this novel technique in this domain that achieves rapid tag identification with high assurance using the minimal number of signal types, which can reduce the decision time and computational burden. This framework introduces three main metrics to measure the system's performance: adaptive identification accuracy ($A$), expected prediction set size $\mu_s$, and expected steps per prediction $\mu_t$. 

The Conformal Prediction provides the set $\mathcal{S}$ with a probability $1-\alpha$ that the true tag label is in this set. For example, if the true tag ID is 1, and there are two data samples. The model can output a set ${\mathcal S}_1=[1]$ for the first sample and ${\mathcal S}_2=[1,2,3]$ for the second sample. With probability $1-\alpha$, the true IDs are in sets ${\mathcal S}_1$ and ${\mathcal S}_2$. However, since set ${\mathcal S}_1$ has a smaller size than ${\mathcal S}_2$, the model is more confident with the decision on the first data sample. Thus, the size of $\mathcal{S}$ can be used as an indicator of the model's confidence. Different from the decision probability that can be generated by typical ML models, the Conformal Prediction has rigorous proof of the coverage \cite{angelopoulos2024theoreticalfoundationsconformalprediction}. In this paper, we define the cumulative coverage for a sample $x$ as $C(x) = 1/|{\mathcal S}|$, where the denominator denotes the set size. The adaptive identification accuracy measures the percentage of correctly predicted instances in a test dataset, considering both the inclusion of the true label in the prediction set $\mathcal{S}$ and whether the cumulative confidence level meets or exceeds a threshold $\tau$. 

The Conformal Prediction framework for pre-trained models is a statistical technique designed to offer predictions with an adjustable confidence level by dynamically setting the size of prediction sets. This technique leverages multiple pre-trained models to make confident predictions as fast as possible (fewer steps), which it then validates based on a cumulative coverage threshold. For each test sample, the prediction set size is obtained to evaluate the certainty of the model's decision. The threshold specifies the minimum cumulative confidence required for a true prediction to be accepted. By focusing on predictions that reach this cumulative confidence threshold, this technique delivers a more robust prediction accuracy while filtering out low-confidence results.

This approach aims to identify the correct NFC tag label by sequentially evaluating test samples across models, beginning with simpler models and advancing to more complex ones only when necessary. We adopt the following progressive data collection process. First, we only get OneLow response to identify the NFC tag. If the identification certainty using $M_2$ is not sufficient, we will gather OneHigh response and use the TwoTypes $M_5$ model for identification. This process continues until we gather all four types of responses and use FourTypes $M_7$ model. At any step, if the identification certainty is sufficient, we can terminate the identification process and generate a high certainty output.  

Specifically, given an NFC tag, we first collect the one subcarrier low data rate response $x_{OL}$ and input to the model $M_2$, which gives a set of predictions $S_1$ containing possible tag labels. For example, $S_1 = [ s_1, s_2, \dots, s_n ]$ represents a set of prediction outputs by model $M_2$ containing possible true labels. Each label in $S_1$ is assigned a confidence score based on the size of the prediction set: 
\begin{equation}
    c_i =
    \begin{cases} 
        \frac{1}{|S_1|}, & \text{if } s_i \in S_1 \\
        0,& \text{else}
    \end{cases}, \quad \text{for } i = 1, 2, \dots, n_t,
\end{equation}
where $C(x)=[c_1,c_2,\cdots,c_{n_t}]$ denotes the cumulative score vector with $n_t$ elements, $n_t$ is the overall NFC tag number, and $|S_1|$ denotes the set size. We compare these scores with the threshold $\tau$. If the maximum score is greater than or equal to $\tau$, we select that as the predicted label and the NFC tag identification process is terminated. On the contrary, if no label meets the threshold, the reader sends a request to collect one subcarrier high data rate response $x_{OH}$.

Next, we add a second test sample $x_{OH}$ and combined with $x_{OL}$. We pass the concatenated test sample to model $M_5$. Model $M_5$ generates a new prediction set. We then update each element in $C(x)$,
\begin{equation}
\label{equ:cxupdate}
c_i =
    \begin{cases} 
        c_i+\frac{1}{|S_2|}, & \text{if } s_i \in S_2 \\
        c_i,& \text{else}
    \end{cases}, \quad \text{for } i = 1, 2, \dots, n_t,
\end{equation}
where $|S_2|$ denotes the set size generated by $M_5$. Similarly, if the maximum score in $C(x)$ is higher than the threshold $\tau$, the identification process is terminated, otherwise, the two subcarriers high data rate response is collected. This process continues with models $M_6$ and $M_7$, updating cumulative confidence scores at each step. If no label reaches the threshold $\tau$ after evaluating with the final model, the predicted label ($y_{{pred}}$) is chosen as
\begin{equation}
    y_{{pred}} = \arg\max(C(x)).
\end{equation}

The Adaptive Identification Accuracy $A$ is given by:
\begin{equation}
    A = \frac{1}{m} \sum_{i=1}^{m} \mathbb{I}({y}_{pred} = Y_{true}),
\end{equation}
where \( m \) is the total number of samples in the test set \( X_{\text{test}} \), and \( \mathbb{I}(\cdot) \) is an indicator function that returns 1  if the true label \( Y_{{true}} \) is same as the predicted tag label $y_{pred}$.
    
The expected prediction set size offers insight into the model's confidence by calculating the average size of the prediction set $S$ across all test samples. Smaller sets imply higher confidence in the model, as a smaller data volume is needed to capture the true label with the required confidence threshold $\tau$. This can be given by:
 \begin{equation}
 \vspace{-1mm}
    \mu_s = \mathbb{E}[|{\mathcal S}|] = \frac{1}{m} \sum_{x \in X_{\text{test}}} |{\mathcal S}(x)|,
\end{equation}
where \( |{\mathcal S}(x)| \) is the size of the prediction set for sample \( x \).

The expected steps per prediction \( \mu_t\) measures the computational efficiency of the Conformal Prediction process by calculating the average number of models required to evaluate each test sample until the cumulative coverage \( C(x) \geq \tau \) condition is met. This metric provides insight into how quickly the model can achieve reliable predictions, balancing between speed and prediction confidence, which is
\begin{equation}
\vspace{-1mm}
    \mu_t=\mathbb{E}[\text{steps}] = \frac{1}{m} \sum_{x \in X_{\text{test}}} f_k(x),
\end{equation}
where \( f_k(x) \) denotes the number of models evaluated for sample \( x \) until the cumulative coverage \( C(x) \geq \tau \) is achieved. A lower expected step count indicates a more efficient model that requires fewer evaluations to reach a confident prediction, while a higher count suggests the model must utilize multiple pre-trained ML/DL models to accumulate sufficient confidence for identification. To provide a clear overview of the Conformal Prediction framework, we summarize all the steps in Algorithm 1.

\begin{algorithm}[t]
\caption{Conformal Prediction-Based NFC Tag Identification}
\begin{algorithmic}[1]
\Require Pre-trained models $\{M_1, M_2, \dots, M_7\}$, test dataset $X_{\text{test}}$ with true labels $Y_{\text{true}}$, confidence threshold $\tau$, risk level $\alpha$, number of samples $m$
\State Initialize metrics: ${\hat a} \gets 0$, $total\_set\_size \gets 0$, $total\_steps \gets 0$

\For{each sample $x \in X_{\text{test}}$}
    \State Initialize prediction set: $S \gets \emptyset$, cumulative confidence $C(x) \gets 0$, step counter $steps \gets 0$
    
    \For{each model $M_i \in \{M_2, M_5, M_6, M_7\}$}
        \State $steps \gets steps + 1$
        \State Obtain prediction set $S$ from $M_i$
        \State Update $C(x)$ using Equ.~\eqref{equ:cxupdate}
        \State $y_{{pred}} = \arg\max(C(x))$
        \If{$\max(C(x_i)) \geq \tau$}
            \State \textbf{break}
        \EndIf
    \EndFor

    \If{$y_{{pred}} = Y_{{true}}[x_{test}]$}
        \State ${\hat a} \gets {\hat a} + 1$
    \EndIf
    \State $total\_set\_size \gets total\_set\_size + |S|$
    \State $total\_steps \gets total\_steps + steps$
\EndFor

\State $A \gets \frac{{\hat a}}{m},~ \mu_s \gets \frac{total\_set\_size}{m},~\mu_t \gets \frac{total\_steps}{m}
$
\State \Return $A, \mu_s, \mu_t$
\end{algorithmic}
\end{algorithm}

\begin{table*}[t]
    \centering
    \captionsetup{justification=centering, font=small, skip=0pt}
    \caption{\scshape \\Metric Comparison of Models on Our Dataset Combinations} 
    \begin{tabular}{@{}cccccccc|cccc@{}}
        \specialrule{.1em}{0em}{0em} 
        \multicolumn{1}{c}{\textbf{Model}} & \multicolumn{8}{c}{\textbf{Test Accuracy(\%)}} & \textbf{CP-adaptive} & \textbf{}  \\ 
        \cmidrule(lr){1-1} \cmidrule(lr){2-8} \cmidrule(lr){9-12}
        \textbf{Name} & \textbf{OneHigh} & \textbf{OneLow} & \textbf{TwoHigh} & \textbf{TwoLow} & \textbf{TwoTypes} & \textbf{ThreeTypes} & \textbf{FourTypes} & \textbf{$A$(\%)} & \textbf{\( \mu_t \)} & \textbf{\( \mu_s \)} & \\ 
        \hline
        \textbf{Random Forest} & 80.92 & 83.52 & 74.94 & 75.31 & 84.96 & 85.21 & 84.96 & 85.81 & 2.63 & 3.28  \\ 
        \textbf{Advanced CNN} & 85.58 & 94.64 & 82.40 & 88.16 & 94.98 & 95.67 & 95.78 &  95.97   
& 1.14 & 1.21
\\ 
        \specialrule{.1em}{0em}{0em} 
    \end{tabular}
    \label{tab:example2}
    \begin{flushleft} 
    {\footnotesize CP-adaptive is the proposed adaptive Conformal Prediction approach. }
    \end{flushleft}
    \vspace{1pt}
\end{table*}

\begin{figure*}[t]
    \centering
    \begin{subfigure}[b]{0.42\textwidth}
        \centering
        \includegraphics[width=1\textwidth]{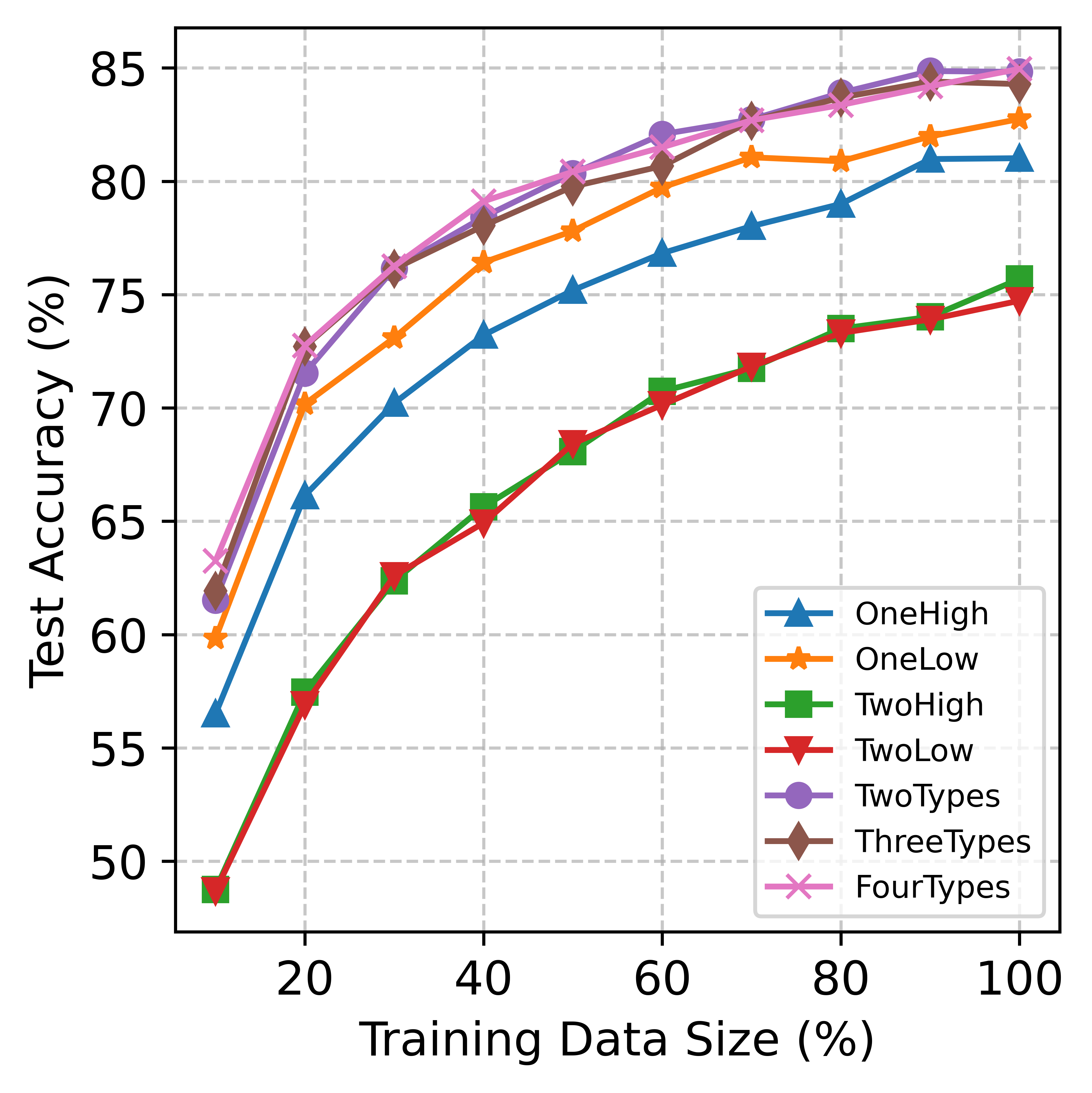}  
        \caption{}
        \label{fig:cnn_cm}
    \end{subfigure}
    \hfill
    \begin{subfigure}[b]{0.42\textwidth}
        \centering
        \includegraphics[width=1\textwidth]{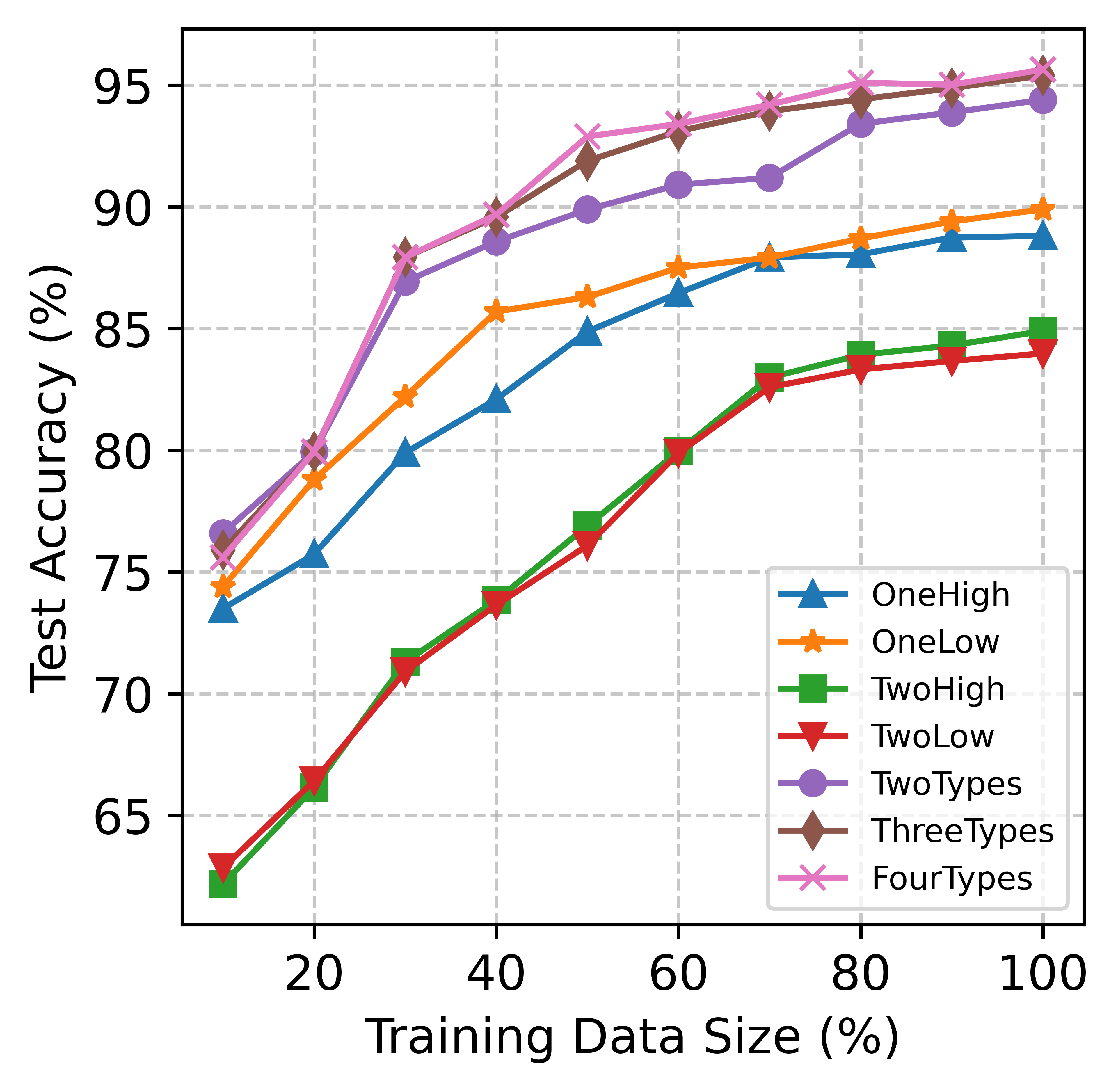} 
        \caption{}
        \label{fig:rf_cm}
    \end{subfigure}
    \caption{Visualization of the impact of data size on (a) Random Forest models performance  and (b) Advanced CNN models performance.}
    \label{fig:accuracy vs datasize_dl}
\end{figure*}

\section{Experiment Analysis and Discussion}
In this section, we evaluate the performance of the Random Forest model and the advanced CNN model using various combinations of the four signal types. Then, we evaluate the adaptive identification using Conformal Prediction.

\subsection{Random Forest Model}
First, we discuss the training accuracy. For the single-subcarrier models, OneHigh achieved an average trained accuracy of $78.21\%$ with no standard deviation across cross-validation folds, indicating stable performance, while OneLow slightly outperformed with an accuracy of $80.11\% \pm 0.10\%$. This suggests that the lower data rate provided a more consistent classification outcome for single-subcarrier NFC tag signals. The two-subcarrier models (TwoHigh and TwoLow) also achieved an average trained accuracy of 72.50\% and 74.26\%  minimal variability ($\pm0.01\%$) and ($\pm0.10\%$) respectively. In contrast, the combination models highlight the advantages of integrating diverse features. The TwoTypes configuration, which merges the OneHigh and OneLow models, achieved the highest recorded accuracy of  $82.32\%$ with zero variability. This result indicates that combining data from different single-subcarrier data rates enhances feature diversity, enabling the classifier to more effectively distinguish NFC tag responses. The more complex ThreeTypes and FourTypes models, which incorporate both single- and two-subcarrier responses, maintained the same high accuracy of 82.21\%. However, the negligible variability observed ($0.00\%$ for ThreeTypes and $0.01\%$ for FourTypes) suggests that further increasing the complexity by adding additional subcarrier types does not improve classification accuracy.

\begin{figure*}[t]
    \centering
    \begin{subfigure}[b]{0.24\textwidth}
        \centering
        \includegraphics[width=\textwidth]{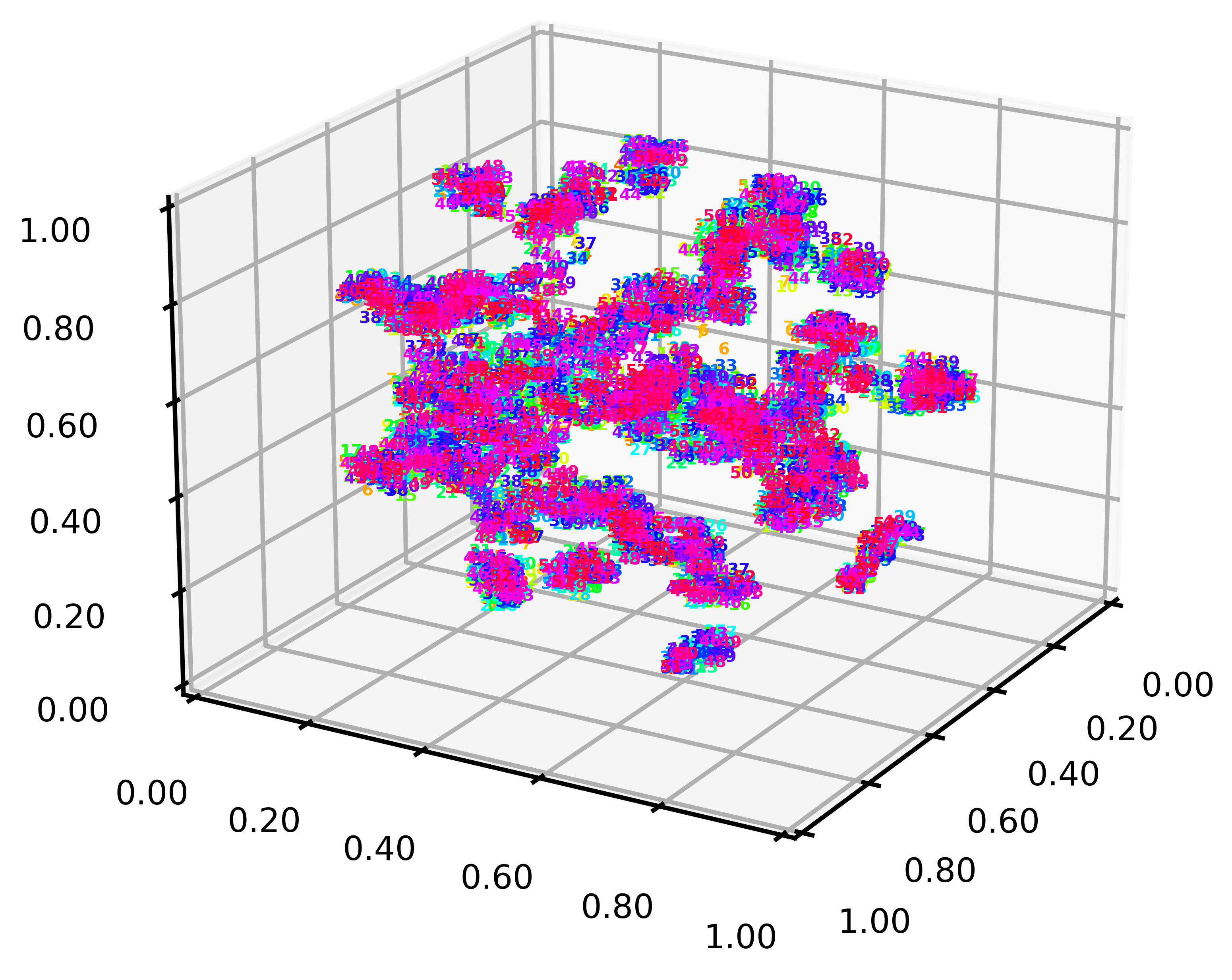}
        \caption{OneHigh}
    \end{subfigure}
    \hfill
    \begin{subfigure}[b]{0.24\textwidth}
        \centering
        \includegraphics[width=\textwidth]{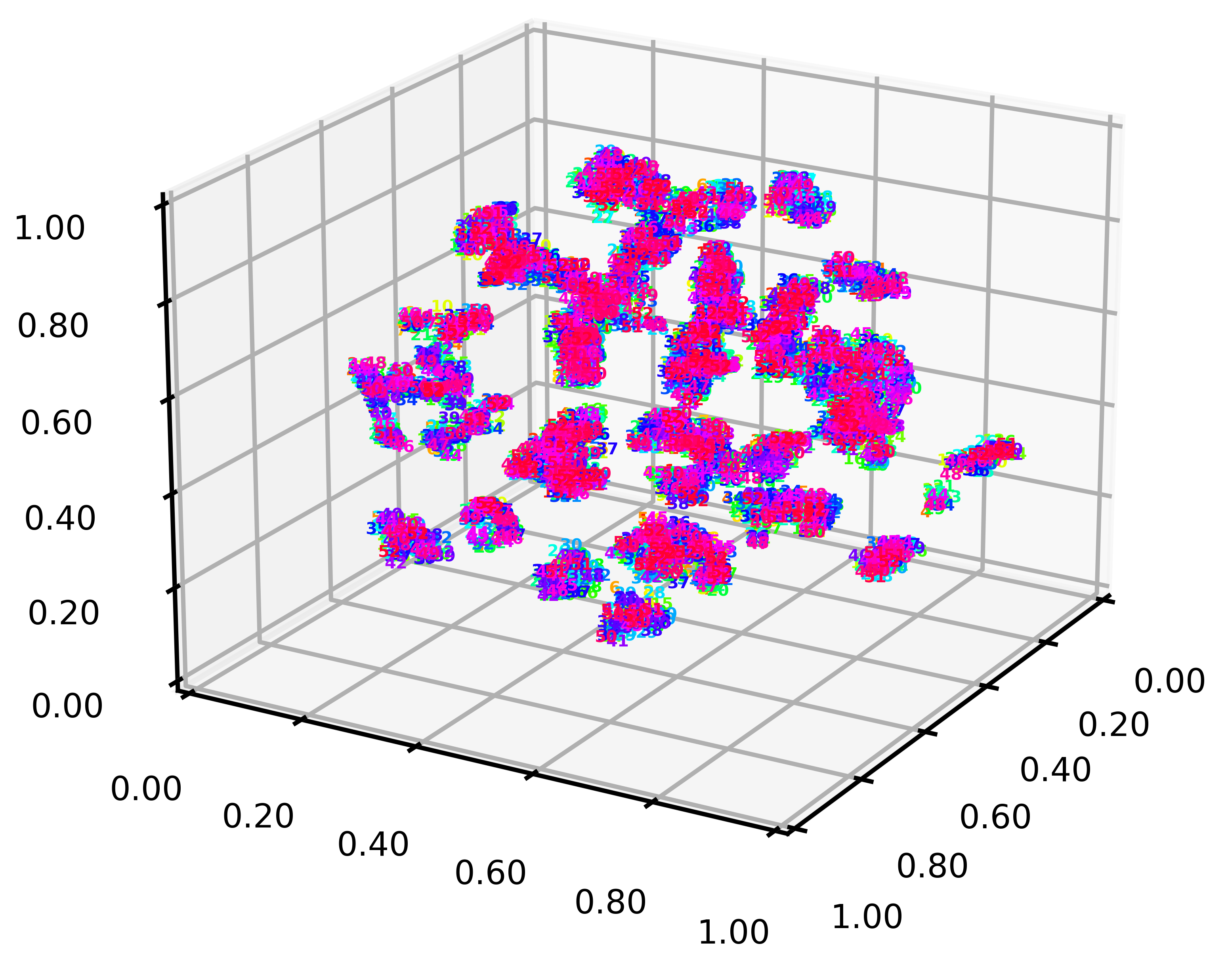}
        \caption{OneLow}
    \end{subfigure}
    \hfill
    \begin{subfigure}[b]{0.24\textwidth}
        \centering
        \includegraphics[width=\textwidth]{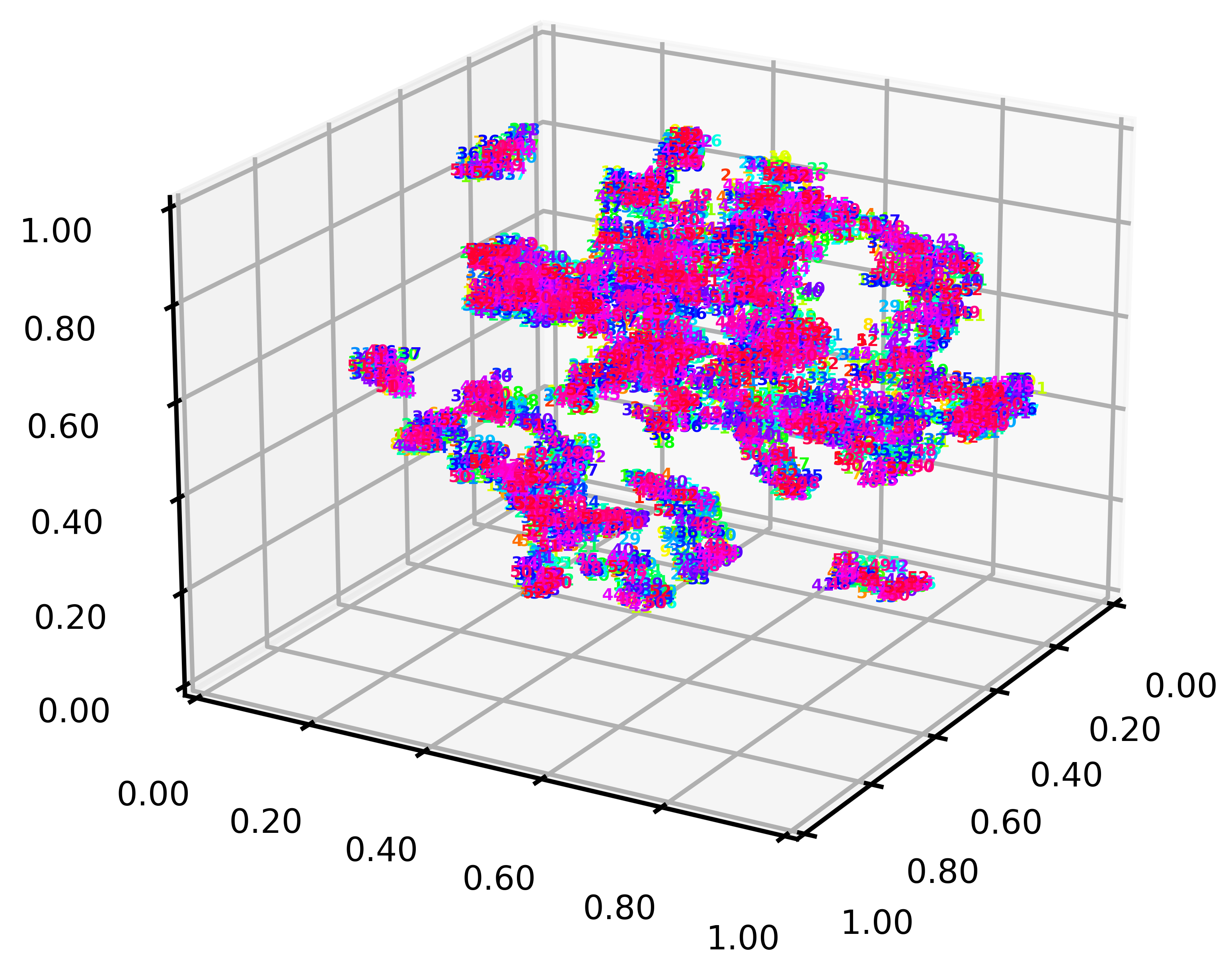}
        \caption{TwoHigh}
    \end{subfigure}
    \hfill
    \begin{subfigure}[b]{0.24\textwidth}
        \centering
        \includegraphics[width=\textwidth]{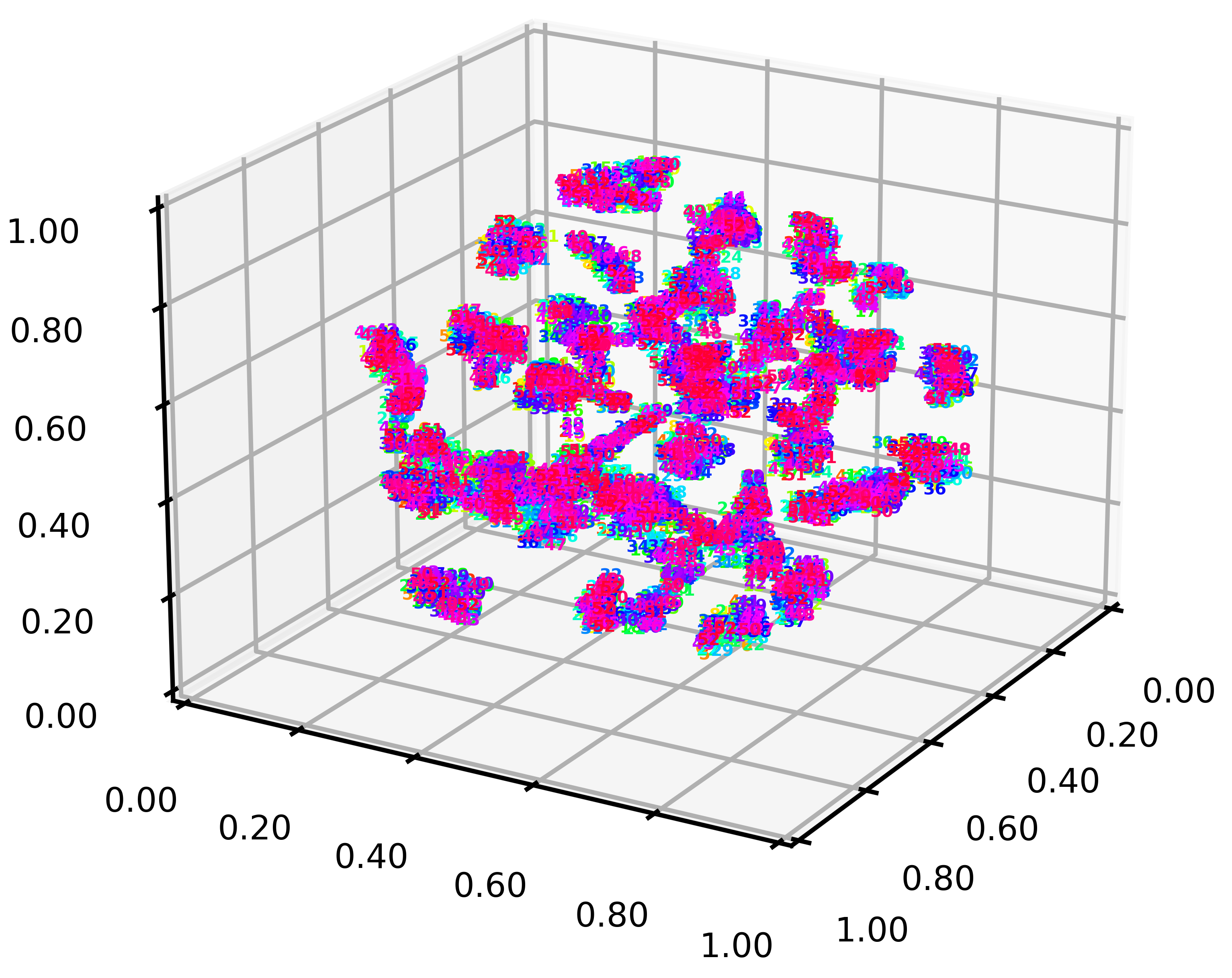}
        \caption{TwoLow}
    \end{subfigure}
    \caption{RF Fingerprints extracted by advanced CNN model, where the output feature maps from the final convolutional layer are projected into a 3-dimensional space using t-SNE. Each color represents a distinct tag label, visually distinguishing different NFC tags.}
    \label{fig:tsne_comparison}
\end{figure*}

\begin{figure*}[t]
    \centering
    \begin{subfigure}[b]{0.32\textwidth}
        \centering
        \includegraphics[width=\textwidth]{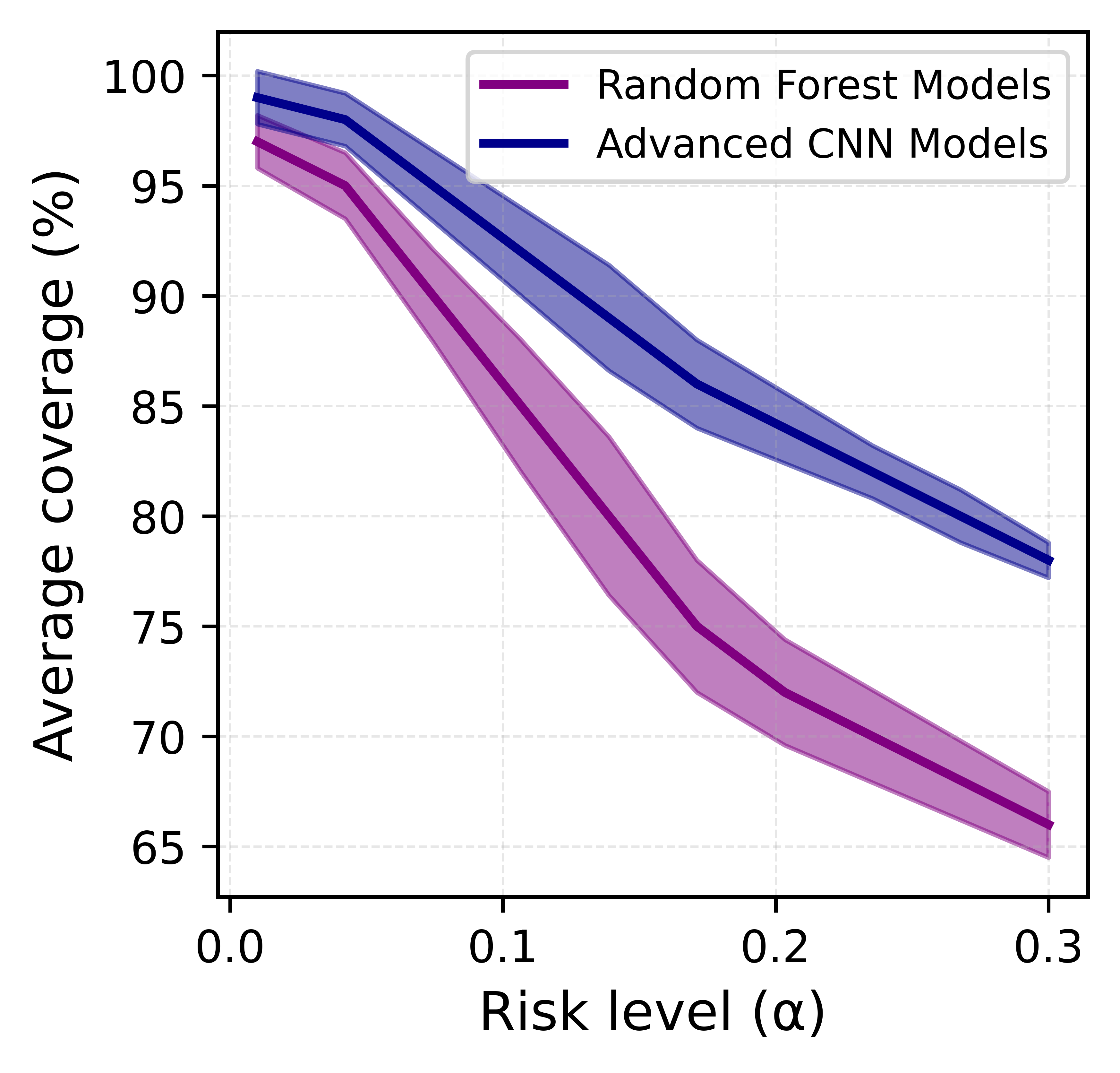}  
        \caption{}
        \label{fig:COVERAGE_RF_CNN_conk}
    \end{subfigure}
    \begin{subfigure}[b]{0.32\textwidth}
        \centering
        \includegraphics[width=0.96\textwidth]{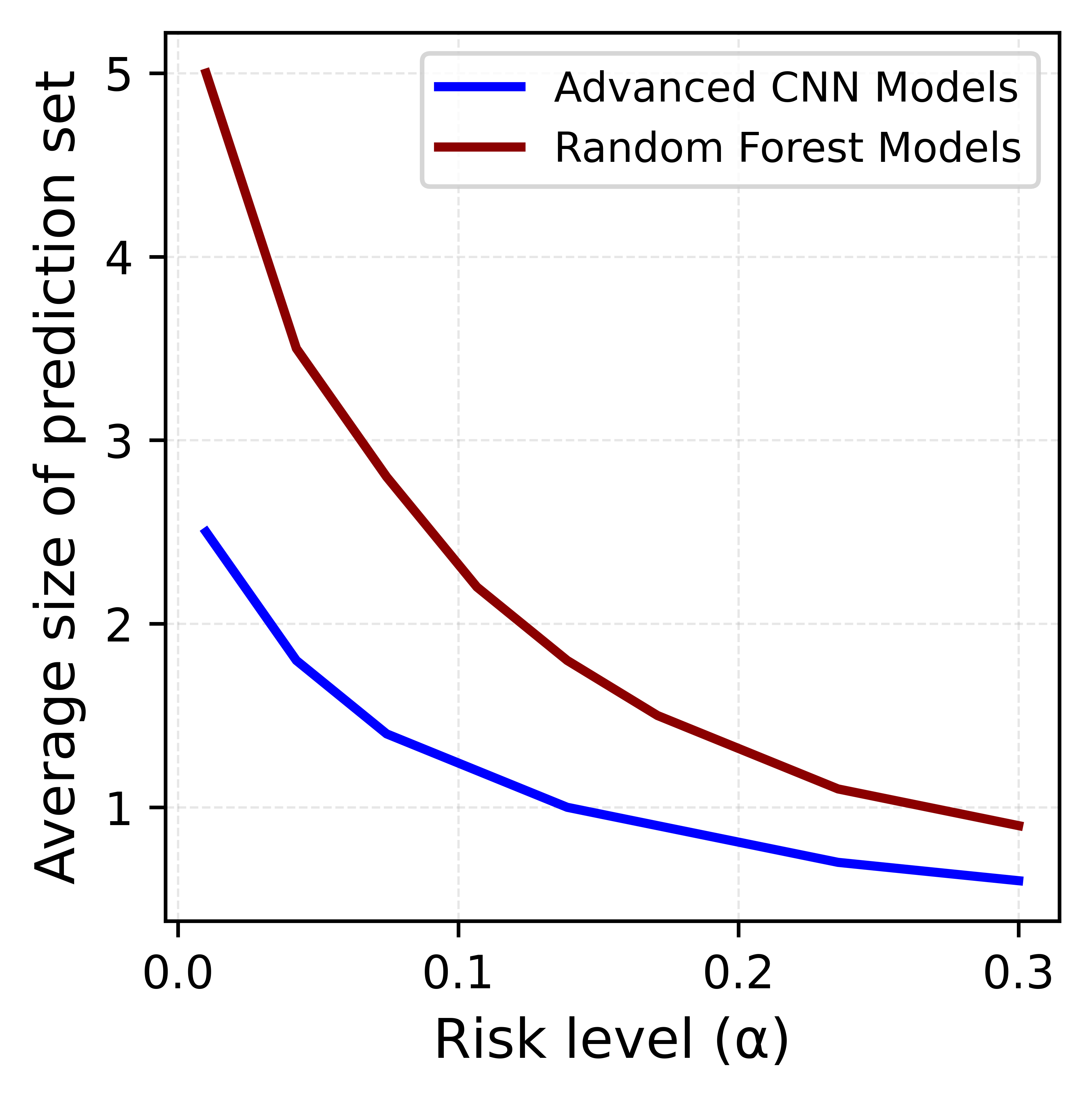} 
        \caption{}
        \label{fig:SET_RF_CNN_conk}
    \end{subfigure}
    \begin{subfigure}[b]{0.32\textwidth}
        \centering
        \includegraphics[width=1.05\textwidth]{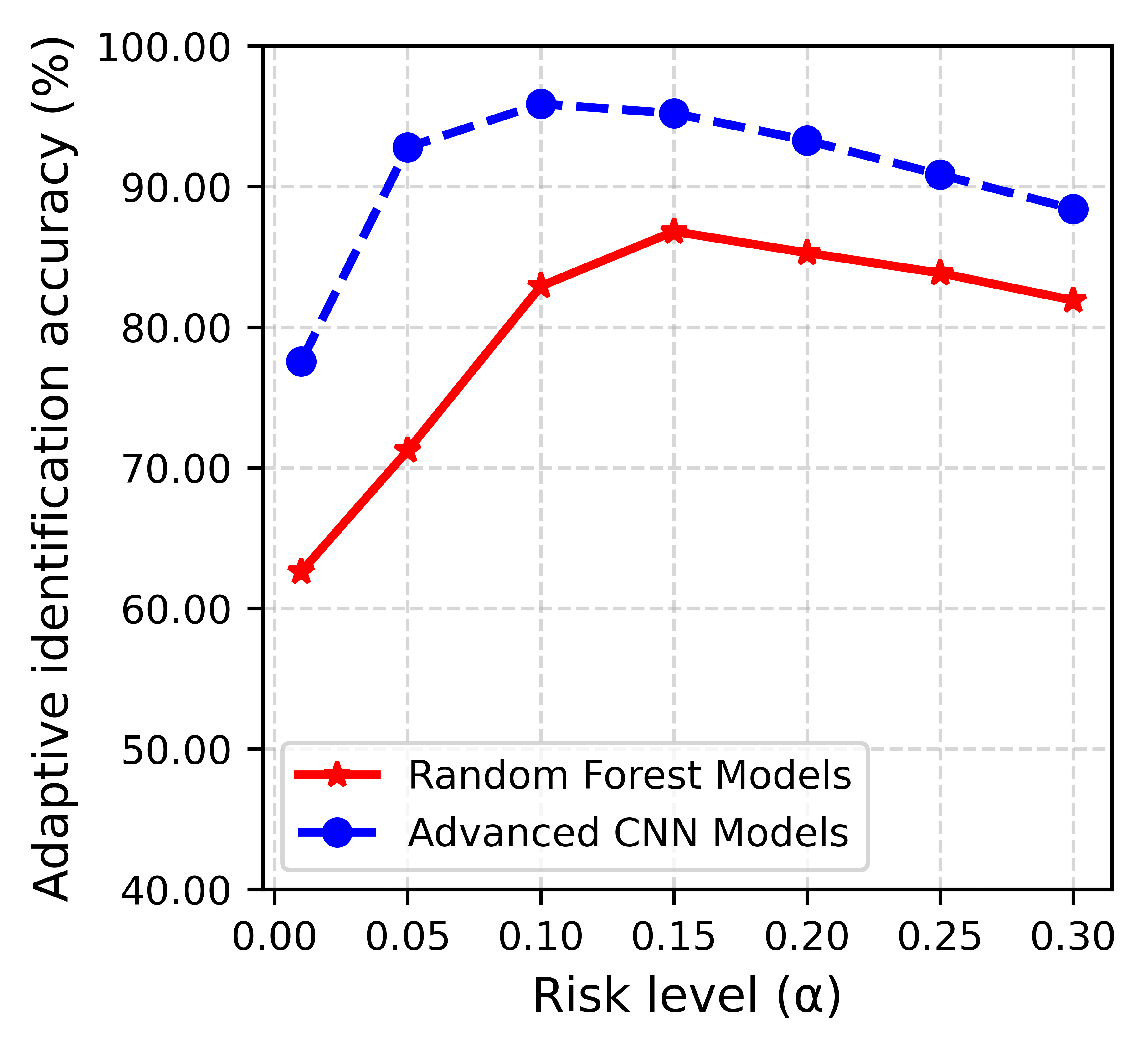} 
        \caption{}
        \label{fig:ACC_RF_CNN_conk}
    \end{subfigure}
    \caption{Comparison of (a) average coverage (proportion of prediction sets containing the true label), (b) average prediction set size, and (c) adaptive identification accuracy across risk levels ($\alpha$) for calibrated Random Forest and Advanced CNN models.}
    \label{fig:covsetacc}
\end{figure*}

Next, we evaluated the trained models on the testing datasets. The accuracy results for various NFC dataset configurations in Table \ref{tab:example2} highlight the impact of subcarrier types, data rates, and their combinations on model performance and stability. The OneHigh model achieves an average accuracy of 80.92\%, the OneLow model shows consistency with its unique features and achieved an average test accuracy of 83.52\%. Models trained on dual subcarrier configurations, such as TwoLow and TwoHigh achieved average test accuracies of 75.31\% and 74.94\% respectively. Among the seven trained models, the highest accuracies are achieved with the ThreeTypes model at 85.21\%, closely followed by TwoTypes and FourTypes with accuracies of 84.96\% and 84.96\%, respectively. The result shows that the one subcarrier signals have more distinct features and the use of multi-channel multi-rate can achieve reliable and accurate identification results.

The ML models exhibit varying levels of accuracy across the training sizes. Fig.~\ref{fig:accuracy vs datasize_dl} demonstrates the relationship between training data size and model accuracy for our seven dataset combinations derived from four primary NFC response types. Initially, increasing the training data size of all 7 dataset combinations significantly improves accuracy, as the model benefits from more representative features. Models combinations incorporating multiple response types tend to perform better, indicating improved generalization. However, in general as the dataset size increases, the accuracy increase margin becomes small, suggesting a saturation point where additional data provides minimal gains. This aligns with the principle of diminishing returns in ML, highlighting that beyond a certain threshold, collecting more data does not significantly enhance performance \cite{9563954}. This insight helps determine the optimal dataset size for NFC card identification without sacrificing accuracy while ensuring computation efficiency.

\subsection{Advanced CNN Model}
A summary of the advanced CNN model's performances is shown in Table \ref{tab:example2}. The model performs exceptionally well on specific data types such as OneLow and comprehensive combinations such as TwoTypes, ThreeTypes and FourTypes. The OneLow dataset alone achieved a high average test accuracy of $94.64\% \pm 0.74\%$, suggesting that its simpler patterns can be learned effectively by the model. Additionally, data types like OneHigh, TwoHigh and TwoLow also yield better results than the Random Forest model, with average test accuracies across 3 training trials of $85.58\% \pm 0.98\%$, $82.40\% \pm 0.18\%$ and $88.18\% \pm 1.02\%$, respectively. Combining multiple signal types generally enhanced model robustness, with the FourTypes combination achieving an average test accuracy of $95.78\% \pm 0.96\%$ by leveraging the diversity of inputs. Similarly, the TwoTypes and ThreeTypes achieved $94.98\% \pm 1.06\%$ and $95.67\% \pm 0.85\%$ respectively. In Fig. \ref{fig:tsne_comparison}, the output features from the final convolutional layer of the advanced CNN model are projected into a 3-dimensional space using t-SNE. It clearly shows that the OneLow model contains well-separated clusters, while models using OneHigh and TwoHigh contain mixed clusters. The trained models of each signal type and their combinations are subsequently saved for adaptive Conformal Prediction.

\subsection{Conformal Prediction for Tag Identification}

\begin{table*}[t]
    \centering
    \captionsetup{justification=centering, font=small, skip=0pt}
    \caption{\scshape \\Summary of trade-offs between coverage, set size, and adaptive identification accuracy for calibrated Random forest and Advanced CNN models for different $\alpha$ values}
    \renewcommand{\arraystretch}{1.2} 
    \setlength{\tabcolsep}{10pt} 
    \begin{tabular}{lccc}
        \toprule
        \textbf{Factor} & \textbf{Low $\alpha$ (~0.01)} & \textbf{Optimal $\alpha$ (~0.05 - 0.15)} & \textbf{High $\alpha$ (~0.30)} \\
        \midrule
        \textbf{Adaptive identification accuracy (Random Forest)} &  62.69\% &  86.81\% &  80.01\% \\
        \textbf{Adaptive identification accuracy (Advanced CNN)} & 74.56\% & 95.89\% &  87.41\% \\
        \textbf{Average prediction set size (Random Forest)} &  5  &  2-3  &  1  \\
        \textbf{Average prediction set size (Advanced CNN)} &  2.5  &  1.6-2.2  &  1  \\
        \textbf{Average coverage (Random Forest)} &   95.00\% &  85.00\% &  65.00\% \\
        \textbf{Average coverage (Advanced CNN)} &  98.00\% & 93.00\% & 82.50\% \\
        \bottomrule
    \end{tabular}
    \label{tab:cov_set_acc}
\end{table*}

\begin{figure*}[t]
    \centering
    \begin{subfigure}[b]{0.42\textwidth}
        \centering
        \includegraphics[width=1\textwidth]{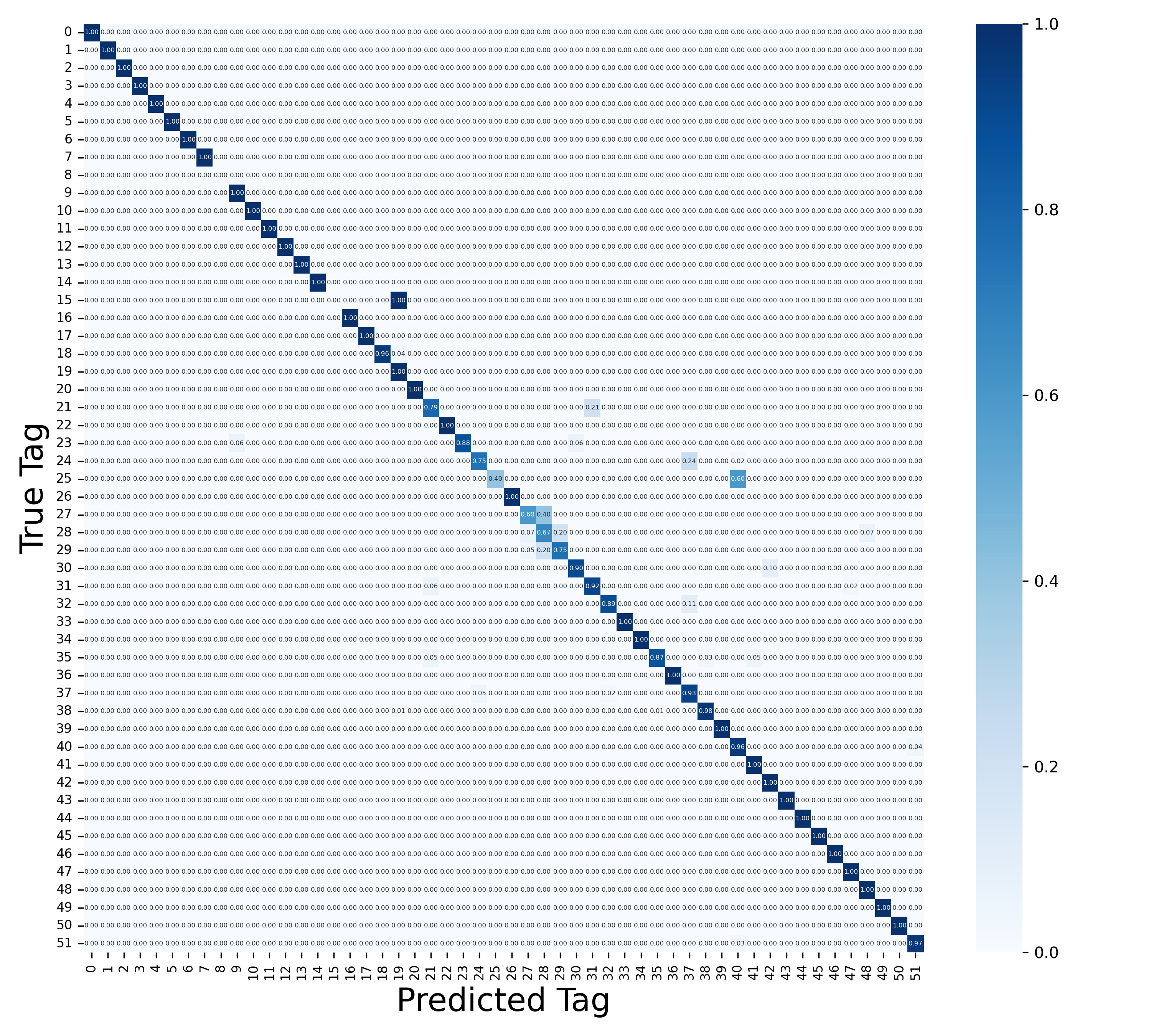} 
        \caption{}
        \label{fig:rf_cm}
    \end{subfigure}
    \hfill
     \begin{subfigure}[b]{0.42\textwidth}
        \centering
        \includegraphics[width=1\textwidth]{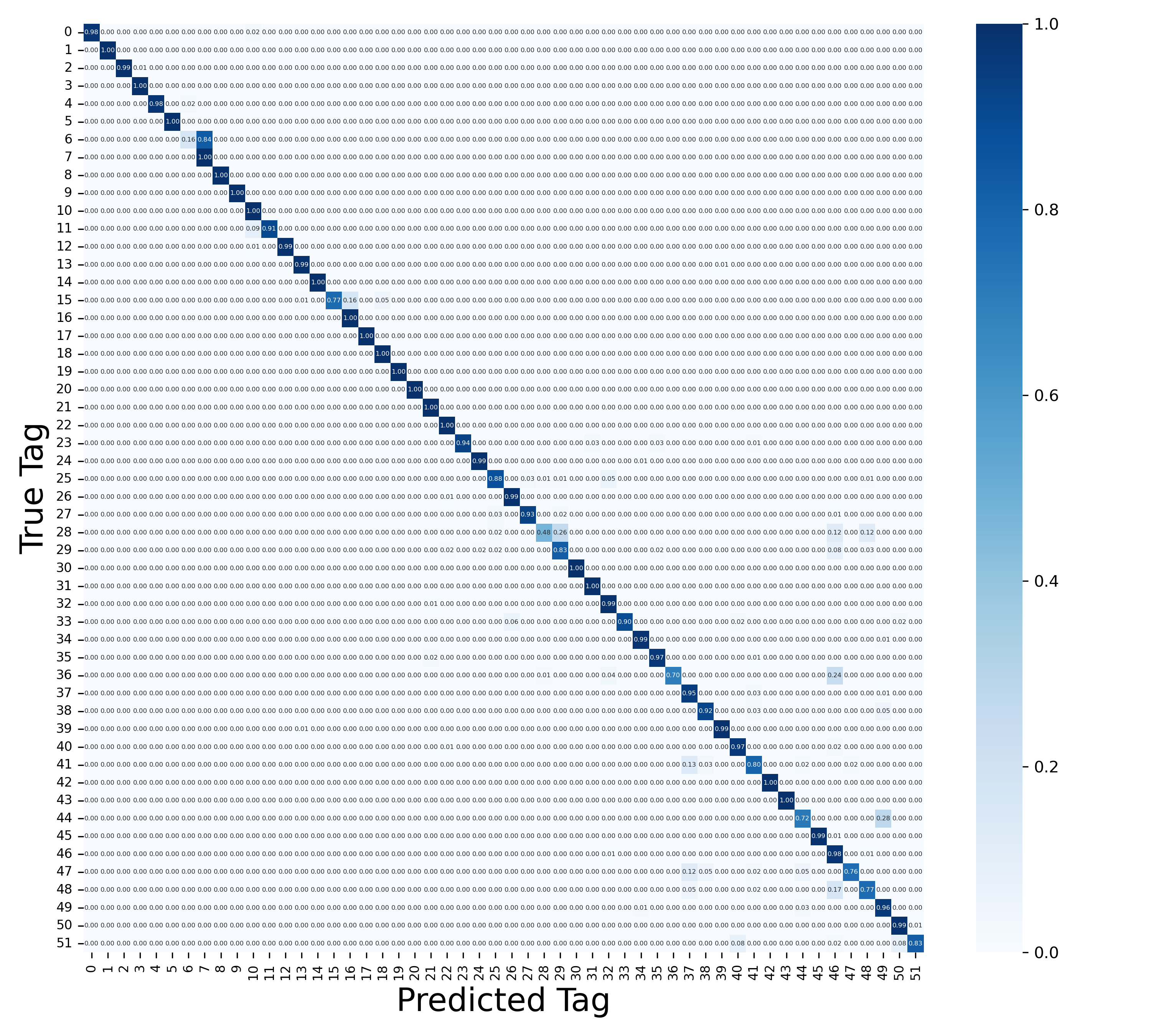}  
        \caption{}
        \label{fig:cnn_cm}
    \end{subfigure}
    \caption{Confusion Matrix illustrating the Performance of the Conformal Prediction Framework Using (a) the Random Forest Model and (b) the Advanced CNN Model.}
    \label{fig:cp_cm_comparison}
\end{figure*}

In Fig. \ref{fig:covsetacc}, the stability and efficiency of the calibrated Random Forest and advanced CNN models using Conformal Prediction across multiple risk levels $(\alpha)$ are shown. Besides accuracy and prediction set size, we also evaluate the average coverage which is the proportion of prediction sets containing the true label. From Fig. \ref{fig:COVERAGE_RF_CNN_conk} both Random Forest and advanced CNN models show consistent decrease of average coverage as $\alpha$ increases. This is because a higher $\alpha$ allows the method to exclude more uncertain predictions, leading to a greater probability of missing the correct label. For an ideally calibrated model, an average empirical coverage should meet the probability $1-\alpha$ (e.g. 90.00\% coverage at $\alpha = 0.10$). The advanced CNN model shows tighter variance compared to the Random Forest model and shows better stability as evidenced in Fig. \ref{fig:COVERAGE_RF_CNN_conk}. As $\alpha$ increases, the expected average prediction set size decreases, aligning with expectations since higher $\alpha$ values allow for more selective prediction sets. Figure \ref{fig:SET_RF_CNN_conk} confirms this observation, showing a clear reduction in set size with increasing $\alpha$. Additionally, advanced CNN model produces smaller prediction sets than the Random Forest model to maintain similar levels of coverage. Similarly, Fig. \ref{fig:ACC_RF_CNN_conk} reveals that both models follow an adaptive accuracy trend with varying $\alpha$. At lower $\alpha$ values, accuracy remains moderate due to larger prediction sets containing more correct labels for each test sample but also more uncertainty. The accuracy peaks at an optimal $\alpha=[0.05-0.15]$, where the models achieve the best balance between adaptive identification accuracy and average prediction set sizes. However, at higher $\alpha$ values, accuracy declines as prediction sets become too small, increasing the risk of incorrect classification for each evaluation. These plots show the expected trade-off between coverage, accuracy and prediction set size and the need to carefully select the optimum user-defined $\alpha$ value.

We evaluate the Random Forest and Advanced CNN models using the Conformal Prediction scheme, with the results presented in Table~\ref{tab:cov_set_acc}. To ensure computational efficiency, we implement a strategic sequential order iteration, where the prediction process cycles through models trained on different data combinations. We start with the "OneLow" pre-trained model as the initial evaluation step, then sequentially progress through the TwoTypes, ThreeTypes, and FourTypes models. This approach simulates a dynamic data aggregation process. If a signal with one carrier at low data rates does not provide high certainty in identification, a one-carrier high data rate signal is collected, and the TwoTypes model is applied. This iterative process continues until all four signal responses are gathered. From Table~\ref{tab:cov_set_acc}, we achieve a balance between accuracy, computational efficiency, and confidence by dynamically adjusting the confidence probability and cumulative prediction threshold. For this test, optimal trade-offs were found at \(\{1 - \alpha = 0.95, \tau = 0.90\}\) for advanced CNN model evaluation and \(\{1 - \alpha = 0.85, \tau = 0.80\}\) for Random Forest model evaluation. Table~\ref{tab:example2} shows that the Random Forest model achieved an adaptive identification accuracy of 85.81\%, demonstrating high confidence and efficiency with a small expected set size of \(\mu_s = 3.28\) and an average of 2.63 steps per prediction. This accuracy is higher than any of the 7 models without adaptive data collection. The Advanced CNN model performed even better, reaching an adaptive identification accuracy of 95.97\% with a lower expected set size of \(\mu_s = 1.21\). This indicates that most predictions require only a single model, with small sets occurring in 20\% of cases. The average number of steps per prediction is 1.14, suggesting that earlier models in the sequence are generally confident in their predictions. Similarly, the performance is better than any of the 7 models without adaptive data collection.

After evaluating the trained models using Conformal Prediction, we generate confusion matrices to compare true vs predicted tag labels across all models in the proposed adaptive approach. Figure~\ref{fig:cp_cm_comparison} presents the confusion matrices for (a) advanced CNN and (b) Random Forest models. Both demonstrate strong accuracy, with values concentrated along the diagonal, indicating correct predictions. The advanced CNN model achieves fewer misclassifications, showing slightly better performance. Off-diagonal entries highlight misclassified tags, providing insights for model improvements. The results confirm the Conformal Prediction framework's reliability in enhancing accuracy. The Advanced CNN proves more robust while maintaining efficiency for real-world deployments.

The adaptive Conformal Prediction framework enhances NFC tag identification accuracy with minimal dataset size while maintaining strong confidence levels. Unlike traditional ML methods that require large datasets for generalization, Conformal Prediction optimizes data usage, reducing computational overhead and data collection needs. This approach outperforms conventional identification flows by providing robust, confidence-based predictions, ensuring reliable NFC tag identification.

\section{Conclusion}
This paper presents an NFC tag authentication system using RF fingerprinting and uncertainty quantification to enhance security and efficiency. Utilizing software-defined radios, a reconfigurable NFC communication data collection platform is developed and machine learning models are trained for NFC tag identification. In addition, this paper uses Conformal Prediction to evaluate the uncertainty of the decision. The system achieves high accuracy with minimal data by adapting to the confidence level of the identification process. This method effectively addresses cloning threats in NFC applications by employing machine learning models to discern unique tag signals. The proposed framework confirms the viability of using fewer yet highly informative responses for tag identification while demonstrating robust and fast approaches to ensure the authenticity of NFC cards in security-sensitive and speed-critical applications.


\bibliographystyle{IEEEtran}
\fontsize{8.5pt}{8.5pt}\selectfont
\bibliography{Ref}

\end{document}